\documentclass[sigconf, screen, authorversion]{acmart}

\copyrightyear{2023} 
\acmYear{2023} 
\setcopyright{acmlicensed}
\acmConference[CHI '23]{Proceedings of the 2023 CHI Conference on Human Factors in Computing Systems}{April 23--28, 2023}{Hamburg, Germany}
\acmBooktitle{Proceedings of the 2023 CHI Conference on Human Factors in Computing Systems (CHI '23), April 23--28, 2023, Hamburg, Germany}
\acmPrice{15.00}
\acmDOI{10.1145/3544548.3580954}
\acmISBN{978-1-4503-9421-5/23/04}
\usepackage{acmart-taps}

\usepackage{balance}       %
\usepackage{graphics}      %
\usepackage[T1]{fontenc}   %
\usepackage{booktabs}
\usepackage{textcomp}

\usepackage{microtype}        %
\usepackage{ccicons}          %
\usepackage[utf8]{inputenc} %

\usepackage[nolist]{acronym}
\usepackage{csquotes}
\usepackage{siunitx}
\usepackage{nicefrac}
\usepackage{cuted}
\usepackage{capt-of}
\usepackage[caption=false]{subfig}
\usepackage[super]{nth}

\newcommand{\nquote}[1]{\enquote{\emph{#1}}}
\newcommand{\pquote}[2]{\enquote{\emph{#1}} (P#2)}
\newcommand{\mpquote}[2]{\enquote{\emph{#1}} (#2)}

\newcommand*{\fullref}[1]{\hyperref[{#1}]{\autoref*{#1} (\nameref*{#1})}} %

\newcommand{\anovaCor}[6]{$F_{#1, #2}=#3$, $p#4$, $\epsilon=#5$\subEtaG{#4}{#6}}
\newcommand{\anova}[5]{$F_{#1, #2}=#3$, $p#4$\subEtaG{#4}{#5}}

\newcommand{\subEtaG}[2]{%
 \ifthenelse{\equal{#1}{\string >.05}}
 {}
 {, $\eta_{G}^{2}=#2$}%
}

\newcommand{\subEta}[2]{%
	\ifthenelse{\equal{#1}{\string >.05}}
	{}
	{, $\eta^{2}=#2$}%
}

\newcommand{\valSi}[3]{$\mu = \SI[round-mode=places,round-precision=1]{#1}{#3}$, $\sigma = \SI[round-mode=places,round-precision=1]{#2}{#3}$}
\newcommand{\val}[2]{$\mu = \num[round-mode=places,round-precision=1]{#1}$, $\sigma = \num[round-mode=places,round-precision=1]{#2}$}

\newcommand{\muemm}{\mu_{\scriptscriptstyle EMM}}

\newcommand{\emmCI}[5]{$\muemm{}~=~\num[round-mode=places,round-precision=1]{#1}#5$ [\num[round-mode=places,round-precision=1]{#3}#5, \num[round-mode=places,round-precision=1]{#4}#5]}

\def\gge{$\epsilon$}

\def\ges{$\eta_{G}^{2}$}

\newcommand{\hypothesis}[1]{\textsc{\texorpdfstring{H\textsubscript{#1}}{H#1}}}

\usepackage{xcolor}
\usepackage{color}

\definecolor{bananayellow}{rgb}{1.0, 0.88, 0.21}
\definecolor{carrotorange}{rgb}{0.93, 0.57, 0.13}
\definecolor{bleudefrance}{rgb}{0.19, 0.55, 0.91}
\definecolor{armygreen}{rgb}{0.29, 0.33, 0.13}
\definecolor{cadmiumgreen}{rgb}{0.0, 0.42, 0.24}
\definecolor{antiquewhite}{rgb}{0.98, 0.92, 0.84}
\definecolor{lightgray}{rgb}{0.83, 0.83, 0.83}
\definecolor{indianred}{rgb}{0.8, 0.36, 0.36}
\definecolor{ivory}{rgb}{1.0, 1.0, 0.94}
\definecolor{citrine}{rgb}{0.89, 0.82, 0.04}
\definecolor{applegreen}{rgb}{0.55, 0.71, 0.0}

\usepackage[textsize=scriptsize]{todonotes}
\presetkeys{todonotes}{backgroundcolor=bananayellow,linecolor=lightgray}{}
\tikzset{/tikz/notestyleraw/.append style={rounded corners=3pt, }}

\newcommand{\inlinemarkupEnv}[1]{%
	\color{#1}%
}

\newlength\myheight
\newlength\mydepth
\settototalheight\myheight{Xygp}
\begin{acronym}[UMLX]
  \acro{HMD}{head-mounted display}
  \acro{GUI}{graphical user interface}
  \acro{HUD}{head-up display}
  \acro{TCT}{task-completion time}
  \acro{SVM}{support vector machine}
  \acro{EMM}{estimated marginal mean}
  \acro{AR}{Augmented Reality}
  \acro{VR}{Virtual Reality}
  \acro{RTLX}{Raw Nasa-TLX}
  \acro{WUI}{Walking User Interface}
  \acro{CSCW}{Computer-Supported Cooperative Work}
  \acro{HCI}{Human-Computer Interaction}
  \acro{ABI}{Around-Body Interaction}
  \acro{SLAM}{Simultaneous Location and Mapping}
  \acro{ROM}{range of motion}
  \acro{EMG}{electromyography}
\end{acronym}

\newcommand{\factorToes}{\textsc{toes}}
\newcommand{\toesAll}{all-toes}
\newcommand{\toesHallux}{hallux}
\newcommand{\toesLateral}{lateral-toes}

\newcommand{\factorPosture}{\textsc{posture}}
\newcommand{\standing}{standing}
\newcommand{\sitting}{sitting}

\newcommand{\factorDirection}{\textsc{direction}}
\newcommand{\dirUp}{extension}
\newcommand{\dirDown}{flexion}

\newcommand{\factorScale}{\textsc{scaling}}
\newcommand{\scale}[1]{#1-scale}

\newcommand{\factorTarget}{\textsc{target}}
\newcommand{\target}[2]{\nicefrac{#1}{#2}-target}

\begin{document}

\def\name{TicTacToes}
\def\plaintitle{\name{}: Assessing Toe Movements as an Input Modality}
\def\plainauthor{First Author, Second Author, Third Author,
  Fourth Author, Fifth Author, Sixth Author}
\def\emptyauthor{}
\def\plainkeywords{Toes; Body-Centric Interaction; Input; Foot; Foot-Based Interaction}
\def\plaingeneralterms{Documentation, Standardization}

\title[\name{}]{\plaintitle}

\author{Florian Müller}
\orcid{0000-0002-9621-6214}
\affiliation{%
	\institution{LMU Munich}
	\city{Munich}
	\country{Germany}
}
\email{florian.mueller@um.ifi.lmu.de}

\author{Daniel Schmitt}
\orcid{0000-0003-0308-0416}
\affiliation{%
	\institution{TU Darmstadt}
	\city{Darmstadt}
	\country{Germany}
}
\email{d4niel.schmitt@gmail.com}

\author{Andrii Matviienko}
\orcid{0000-0002-6571-0623}
\affiliation{%
	\institution{KTH Royal Institute of Technology}
	\city{Stockholm}
	\country{Sweden}
}
\email{matviienko.andrii@gmail.com}

\author{Dominik Schön}
\orcid{0000-0003-2704-2852}
\affiliation{%
	\institution{TU Darmstadt}
	\city{Darmstadt}
	\country{Germany}
}
\email{schoen@tk.tu-darmstadt.de}

\author{Sebatian Günther}
\orcid{0000-0003-1281-2180}
\affiliation{%
	\institution{TU Darmstadt}
	\city{Darmstadt}
	\country{Germany}
}
\email{guenther@tk.tu-darmstadt.de}

\author{Thomas Kosch}
\orcid{0000-0001-6300-9035}
\affiliation{%
	\institution{HU Berlin}
	\city{Berlin}
	\country{Germany}
}
\email{thomas.kosch@hu-berlin.de}

\author{Martin Schmitz}
\orcid{0000-0002-7332-3287}
\affiliation{
	\institution{Saarland University,\ Saarland Informatics Campus}
	\city{Saarbr\"{u}cken}
	\country{Germany}
}
\email{mschmitz@cs.uni-saarland.de}

\renewcommand{\shortauthors}{Müller et al.}

\acresetall

\begin{abstract}

From carrying grocery bags to holding onto handles on the bus, there are a variety of situations where one or both hands are busy, hindering the vision of ubiquitous interaction with technology. Voice commands, as a popular hands-free alternative, struggle with ambient noise and privacy issues. As an alternative approach, research explored movements of various body parts (e.g., head, arms) as input modalities, with foot-based techniques proving particularly suitable for hands-free interaction. Whereas previous research only considered the movement of the foot as a whole, in this work, we argue that our toes offer further degrees of freedom that can be leveraged for interaction. To explore the viability of toe-based interaction, we contribute the results of a controlled experiment with 18 participants assessing the impact of five factors on the accuracy, efficiency and user experience of such interfaces. Based on the findings, we provide design recommendations for future toe-based interfaces.
\end{abstract}
\acresetall



\begin{teaserfigure}
	\includegraphics[width=\textwidth]{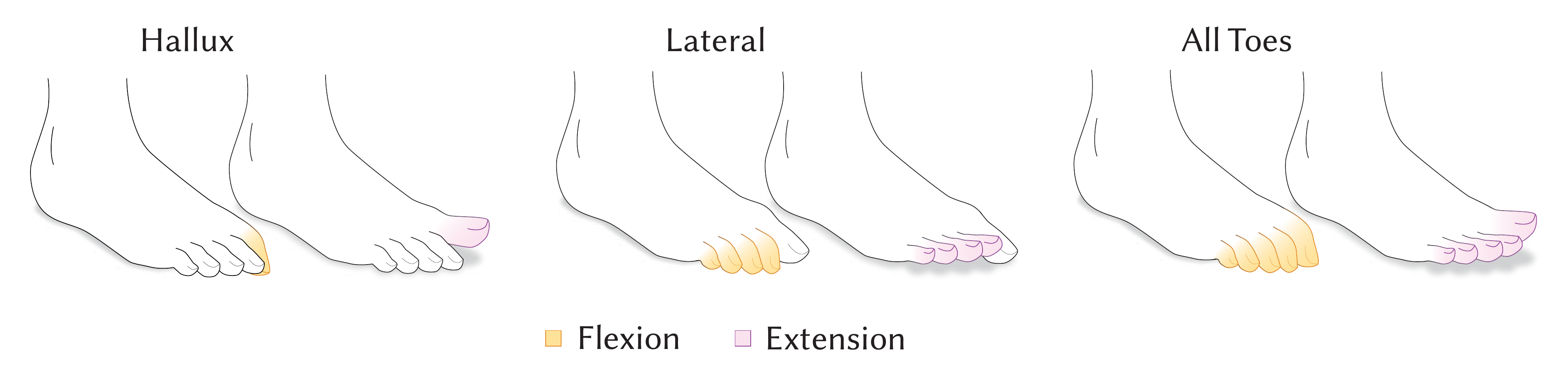}
	\caption{In this work, we investigate the feasibility of a toe-based input modality for interacting with computing systems. For this, we investigate the influence of five factors on the accuracy, efficiency, and user expierience of such interfaces. In addition to the \factorDirection{} and the \factorToes{} as illustrated in this figure, we further investigate the influence of the \factorPosture{}, the \factorScale{} of the interaction range and the location of the \factorTarget{}.}
	\label{fig:teaser}
	\Description{An illustration of human feet. The toes are extended and flexed in the groups all toes, hallux and lateral toes.}
\end{teaserfigure}

\maketitle

\section{Introduction}
\label{sec:tictactoes:intro}

\acresetall
From tapping on a smartphone or touching a tangible user interface~\cite{Ishii2008} to mid-air gestures for \acp{HMD}~\cite{Mistry2009, Colaco2013}, we use our hands to interact with today's technology in most situations. This emphasis on the manual operation of interfaces in \ac{HCI} research has its roots in human evolution and the upright gait that freed the hands from the task of locomotion. Subsequently, we developed a considerable dexterity for fine-grained manipulations and interactions~\cite{Velloso2015} that we employ today in interactions with the physical world and computer systems. However, there are a variety of situations where it is undesirable to use our hands to interact (e.g., when we are cooking something and have dirty hands) or even physically impossible (e.g., when we are carrying things in our hands).

To overcome these challenges, research explored hands-free~\cite{fejtova2009hands} interaction styles. Besides voice-based interaction~\cite{10.1093.iwc.iwz016} which is known to be limited to quiet environments and poses privacy issues~\cite{Koelle2017,Starner2002}, research proposed leveraging the movements of other body parts. Such body-centric~\cite{Wagner2013} interaction techniques have been explored for various body parts such as the eyes~\cite{Lukander2013, Piumsomboon2017}, the head~\cite{Morris2000, Lu2020}, or combinations~\cite{Spakov2012, Gobel2013, Chatterjee2015} of these. Among these, foot-based techniques emerged as particularly suitable for hands-free interactions with desktop PCs~\cite{Silva2009}, smartphones~\cite{Fan2017}, or \acp{HMD}~\cite{Muller2019}. Interestingly, these techniques usually only consider the feet as a whole, ignoring the internal degrees of freedom of the feet: our toes. 

While previous work has only considered the human toes for binary switches~\cite{thorp1998invention, Carrozza2007}, we argue that the degrees of freedom of human toes allow for more fine-grained interaction. We can voluntarily extend and flex our toes, move them independently of each other, and hold them in different positions. Thus, the degrees of freedom of toes might qualify as a simple and readily available input modality for hands-free interaction in a variety of situations. However, to exploit such movements as part of future interaction techniques, we require a thorough understanding of the human capabilities for voluntary toe movements and the factors that influence the accuracy, efficiency, and user experience of such movements.

In this paper, we fill this gap and extend the state of research on body-centric interaction by investigating the human capabilities for toe movement as an input modality. The contribution of this paper is two-fold: First, we present the results of a controlled experiment with 18 participants investigating the influence of the \factorToes{}, the \factorPosture{}, the \factorDirection{}, the \factorScale{}, and the \factorTarget{} on the accuracy, efficiency, and user experience of voluntary toe movements. Second, based on the results of the controlled experiment, we provide a set of guidelines for designing toe-based interaction techniques.

\section{Related Work}
\label{sec:tictactoes:rw}

A large body of related work exists on 1) body-centric interaction techniques that strongly influenced our work. In the following section, we discuss these works, with a particular focus on foot-based interaction techniques. Further, we present the basics of 2) biomechanics of toe movements.

\subsection{Body-Centric Interaction}

With the spread of (low-cost) sensor hardware and advances in computer vision, research began to investigate movements of the human body as an input modality in the area of body-centric~\cite{Chen2012} (also body-based~\cite{Silva2009}) interaction.

Based on their natural dexterity, research efforts focused on interaction using the hands. As a prominent example, on-body~\cite{Harrison2012} interaction techniques allow for (multi-) touch input on different body parts such as the torso~\cite{Hamdan2017}, hands~\cite{Tamaki2010, Dezfuli2012, Muller2015b}, and arms~\cite{Harrison2010, Weigel2014}. Without visual output, research also explored other locations, such as the ear~\cite{Lissermann2013, Kikuchi2017} or the face~\cite{Yamashita2017, Serrano2014}. Further, research investigated mid-air gesture interfaces around the body~\cite{Muller2020}: Mistry et al. \cite{Mistry2009} proposed a wearable interface, and Colaco et al. \cite{Colaco2013} showed how to capture single-handed gestures. Other examples include proximity-based interfaces~\cite{muller2015a}, single-finger gestures~\cite{Buchmann2004}, or combinations with gaze~\cite{Heo2010, Spakov2012}. While both on-body and mid-air interfaces show several benefits (e.g., the fast and direct manipulation of content), these techniques require at least one, often both hands and, thus, cannot support situations when the user's hands are not free. Further, gestures are prone to fatigue, also known as the \emph{gorilla arm syndrome}~\cite{Hincapie-Ramos2014}, and, thus, not suited for longer interactions.

As another promising direction, research proposed various hands-free body-centric interaction techniques that leverage the degrees of freedom of other body parts, such as the head~\cite{Hueber2020, Yan2018} and the eye~\cite{Mateo2008, Nilsson2009}. Further examples include the ear~\cite{Roddiger2021}, the mouth~\cite{GallegoCascon2019, Sahni2014}, and the face~\cite{Matthies2017}, as well as interpreting blow gestures~\cite{Chen2015, Reyes2016} or combinations of multiple body parts~\cite{Matthies2013a, Fejtova2009}. As the most related area of such hands-free body-centric interfaces, foot-based interaction techniques have a long tradition in the operation of industrial machinery~\cite{barnes1942pedal,Corlett1975,Keoemer1971,Pearson1986, Barnett2009}. Consequently, the HCI community has explored such foot-based interfaces for various use cases. In the following section, we discuss these works.

\subsection{Foot-based Interaction}

\begin{figure*}[ht!]
	\subfloat[Pre-trial screen\label{fig:tictactoes/task:1}]
	{\includegraphics[width=.24\linewidth]{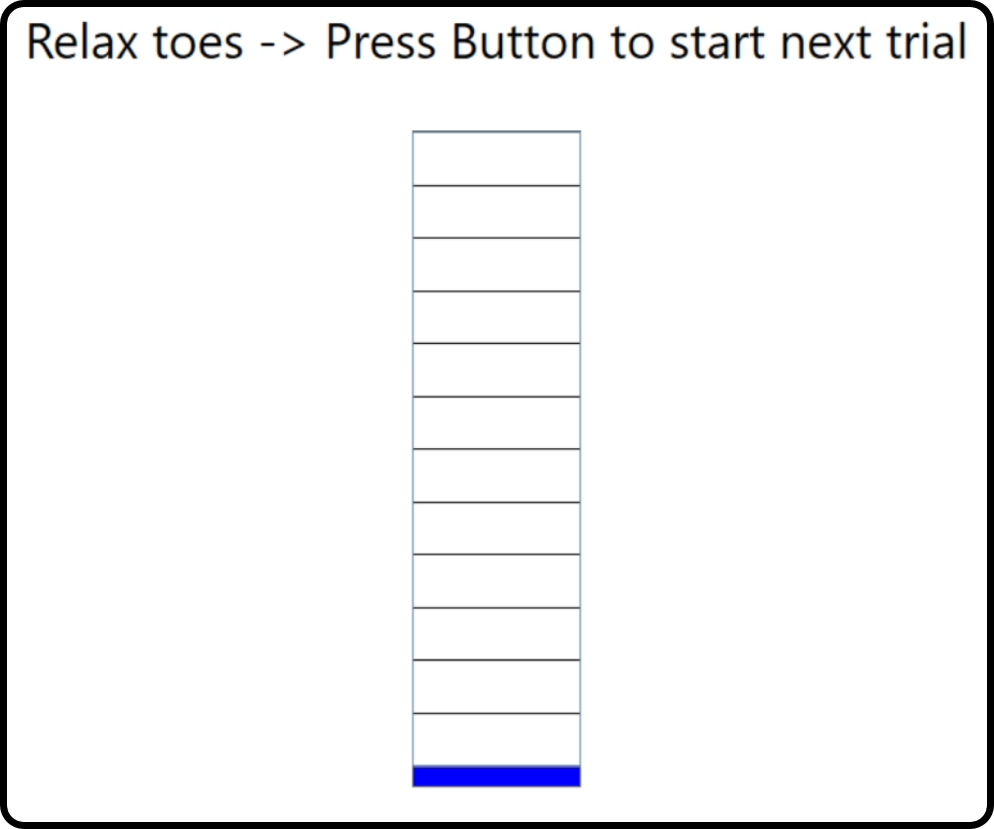}}\hfill
	\subfloat[Task 1: Moving to the target\label{fig:tictactoes/task:2}]
	{\includegraphics[width=.24\linewidth]{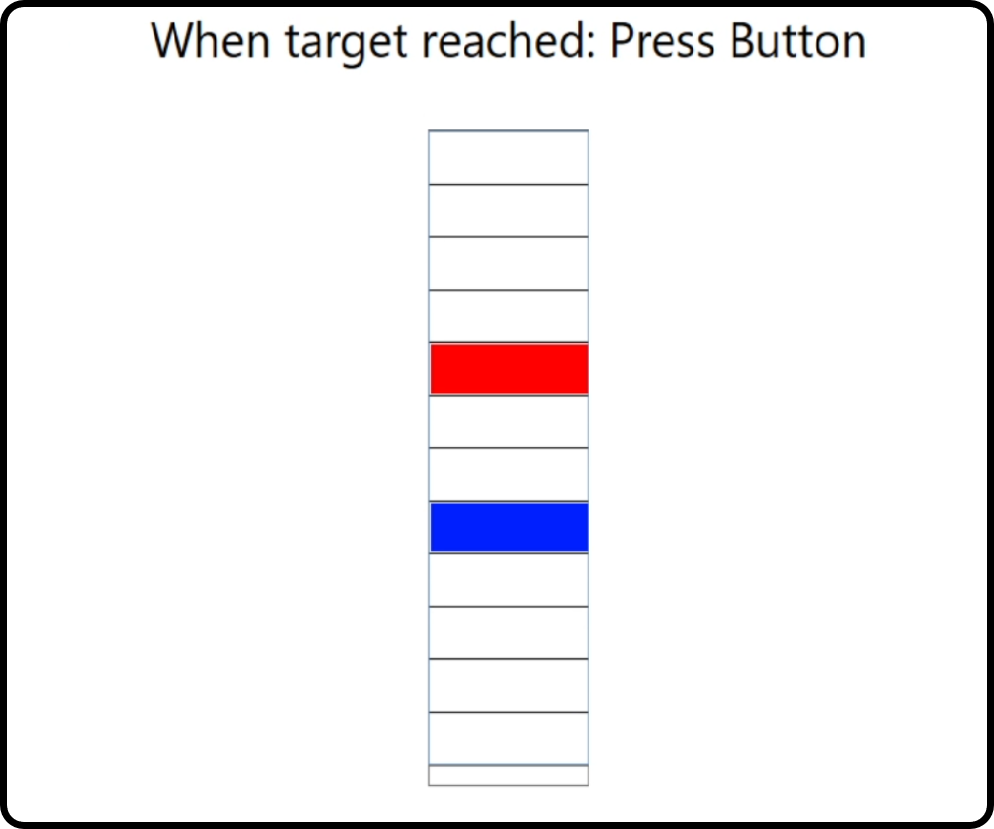}}\hfill
	\subfloat[Task 1: Selecting the target\label{fig:tictactoes/task:3}]
	{\includegraphics[width=.24\linewidth]{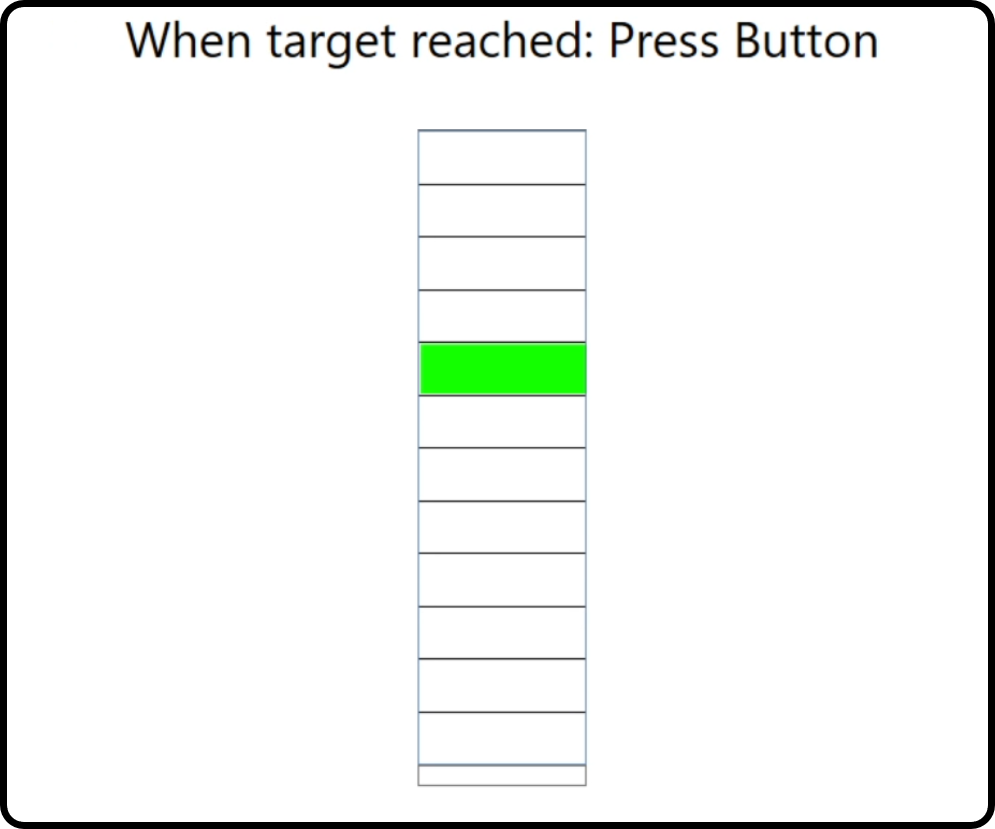}}\hfill
	\subfloat[Task 2: Holding on the target\label{fig:tictactoes/task:4}]
	{\includegraphics[width=.24\linewidth]{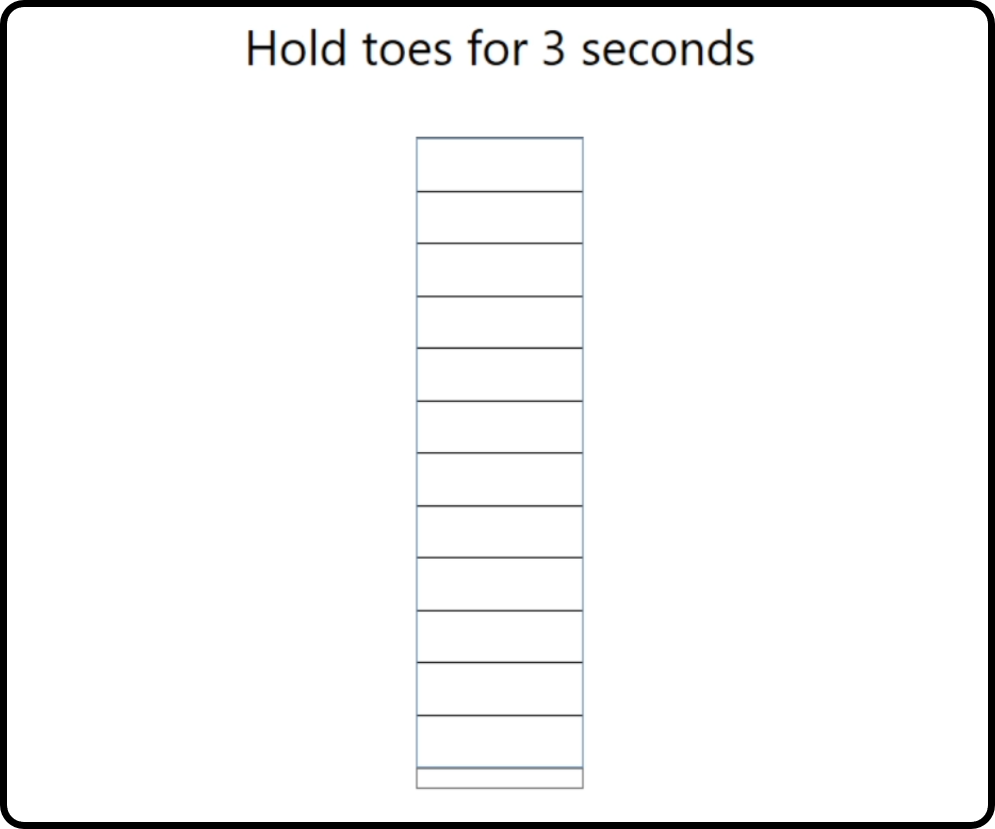}}
\caption{After participants indicated readiness for the next trial by clicking (a), the system highlighted the target (b, red) and visualized the movement of the respective toes on the condition (a, b, blue). While the participant's toe was positioned over the target, the shared cell was highlighted in green (c). After the participant confirmed reaching the target by clicking, the second task started, in which the participants had to keep their toes stationary without visual feedback (d).}
\Description[A 4-part illustration of the task in the experiment.]{An illustration of the task in the experiment. The Figure shows four screens in the order they appeared in each trial. They all contain a vertical scale that is divided into multiple areas with highlightings for the current position of the toes and the target.}
\end{figure*}

Research investigated foot-based interaction techniques in different postures, such as seated~\cite{english1967display, Velloso2015a}, standing~\cite{Saunders2016}, and walking~\cite{Yamamoto2008, Smus2010}. Such foot-based interfaces have been used as input modalities for desktop~\cite{Silva2009} or mobile~\cite{Lv2014a} applications or to operate a smartphone in the user's pocket~\cite{Barnett2009, Fan2017, Han2011}. \citet{Pakkanen2004} compared hand- and foot-based operation of a trackball for spatial tasks and found accuracy and efficiency suitable for secondary tasks. As a more recent trend, research also investigated the applicability of foot-based interfaces for \acp{HMD}. Fukahori et al. \cite{Fukahori2015} leveraged shifting the user's weight on their foot for subtle gestures to control \ac{HMD} interfaces. Saunders et al.~\cite{Saunders2015a, Saunders2016} studied tapping with different parts of the foot, e.g., the heel or toes, and kicking movements for interaction. In addition, \citet{Muller2019} compared foot tapping for direct and indirect interaction. Besides the sole use as an input modality, foot interaction has been used in conjunction with other input modalities. Lv et al.~\cite{Lv2013, Lv2014, Lv2015} explored hand gestures in combination with foot movements such as rotating or tapping, and \citet{Lopes2019} how such a combination can be used for 3D object manipulation. Further, \citet{Muller2020d} showed how the process of locomotion through foot movements can be used for interaction.

Further, \citet{10.1007.978-3-030-78092-0_34} showed that a combined foot and gaze input modality could improve target acquisition. As practical examples, \citet{Gobel2013} and \citet{Klamka2015} combined foot movements through a foot rocker and a foot joystick with gaze input for zooming and panning, and \citet{Rajanna2016} presented a  mouse replacement for desktop environments. Highly related, \citet{Matthies2013} integrated various sensors into the sole of a shoe to sense actions such as walking and jumping. Further, their system allows sensing hallux-presses as a binary input switch through capacitive sensing. Similarly, research explored the use of hallux presses as binary input switches for covert interaction~\cite{thorp1998invention} and for hands-free control of robotic systems~\cite{Carrozza2007}.

While the human feet, in general, and their suitability for interacting with computer systems have been thoroughly investigated, the feasibility of interaction using the toes has not been systematically studied beyond binary single-toe interactions. Such toe-based interactions could provide an additional input modality for subtle and privacy-preserving in-shoe interactions. To explore the feasibility of such an input modality and to provide a foundation for future interaction techniques, in this paper, we contribute an in-depth analysis of the human capabilities for such interactions.

\subsection{Biomechanics of Toe Movements}

Just like hands, human feet have five digits called toes, numbered from 1 to 5 starting from the hallux (also: big toe). Each of the toes consists of three \emph{phalanx bones}, except for the \emph{hallux}, which consists of only two bones~\cite{Drake2020}. Similar to the fingers, the joints of the phalanx bones (\emph{interphalangeal joints}) of the toes allow for four types of movement: (Downward) \emph{flexion} and (upward) \emph{extension} and \emph{abduction}, and \emph{adduction} to spread and return the toes to a resting position, respectively~\cite{Lippert2011}. Although lateral \emph{abduction} and \emph{adduction} movements are possible to a certain extent, the joints are designed primarily as hinge joints for upward and downward movement~\cite{Baehler1986}. Flexion and extension of the toes are mainly driven by the \emph{flexor digitorum brevis} and the \emph{extensor digitorum brevis} muscles, which are attached to all toes except the hallux~\cite{Drake2020}. The hallux is flexed (\emph{flexor hallucis longus} and \emph{flexor hallucis brevis}) and extended (\emph{abductor hallucis} and \emph{adductor hallucis}) by its own set of muscles~\cite{Drake2020}. Consequently, all toes except hallux cannot be moved separately. Thus, in the following, we will consider these two groups separately as the hallux 
(\aptLtoX{1\textsuperscript{st}}{\nth{1}} toe) and the lateral toes (\aptLtoX{2\textsuperscript{nd}-5\textsuperscript{th}}{\nth{2}-\nth{5}} toe).

The toes are essential in human gait while walking or running for stabilization and control~\cite{Bojsen-Moller1979, Rolian2009}. This largely excludes voluntary movements --- i.e., movements not based on reflexes to behavioral stimuli but on an intention to act~\cite{Rizzolatti2013} --- during simultaneous locomotion because they would impair the gait. During other phases, such as standing, sitting, or lying down, however, we can control the discussed degrees of freedom of our toes voluntarily.

Analyses of the \ac{ROM} of the toes showed $\approx$40° for flexion and $\approx$50° for extension for active movements~\cite{Creighton1987} (i.e., movements generated by muscle force as opposed to passive movements induced by an external counterforce) with large variations between individuals~\cite{Joseph1954}. Highly related, Yao et al. \cite{Yao2020} recently presented an evaluation of the movement characteristics of fingers and toes. In their work, the authors compared the efficiency of moving the hallux and \aptLtoX{2\textsuperscript{nd}}{\nth{1}} toe over different distances for flexion and extension movements. However, the authors did not explore the combined interaction with all toes and the influence of body posture. Further, the authors did not focus on such movements' accuracy and user experience, which is a key concern for potential use as an interaction modality. Therefore, the accuracy and efficiency with which the toes can be used to perform targeted movements, not only to complete deflection but also to select intermediate targets, is still an open question. Further, to our knowledge, there is no subjective assessment of such movements in terms of comfort and convenience.

\section{Methodology}
\label{sec:tictactoes:methodology}

\begin{figure*}[ht!]
\includegraphics[width=\linewidth]{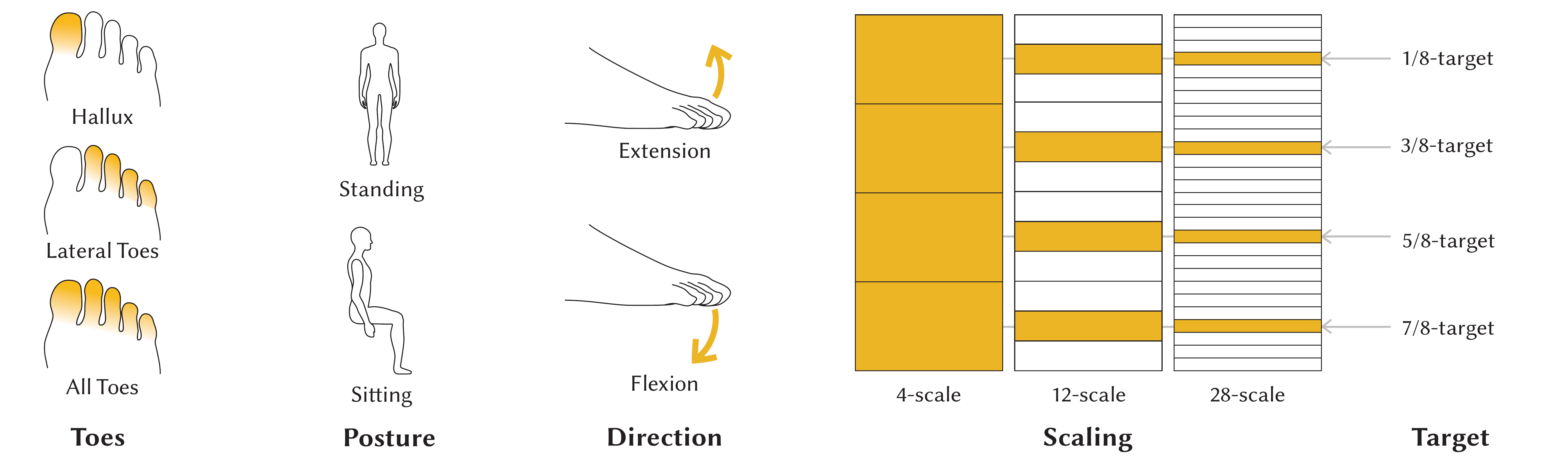}
\caption[]{The independent variables studied in the experiment with their respective levels. From left to right: \factorToes{} (\toesAll{}, \toesHallux{}, \toesLateral{}), \factorPosture{} (\standing{}, \sitting{}), \factorDirection{} (\dirUp{}, \dirDown{}), \factorScale{} (\scale{4}, \scale{12}, \scale{8}), and \factorTarget{} (\aptLtoX[graphics=no, type=html]{1/8-target}{\target{1}{8}}, \aptLtoX[graphics=no, type=html]{3/8-target}{\target{3}{8}}, \aptLtoX[graphics=no, type=html]{5/8-target}{\target{5}{8}}, \aptLtoX[graphics=no, type=html]{7/8-target}{\target{7}{8}}).}
\label{fig:tictactoes/factors}
\Description[An illustration of the five independent variables in the experiment.]{An illustration of the independent variables in the experiment. From left to right, the figure depicts a) human toes with highlightings for all toes, hallux, and lateral toes, b) the posture as standing and sitting, c) the direction of interaction (flexion and extension), d) the different scalings from 4 to 28 subdivisions, e) the four targets from 1/8 to 7/8 used in the experiment.}
\end{figure*}

We conducted a controlled experiment to investigate the accuracy, efficiency, and user experience of toe movements as an input modality for computer systems. The presence of footwear and its type, design, and cut, as well as the interplay of these with different foot shapes, has a very individual effect on the freedom of movement of the toes. This impacts the possible interactions as well as their accuracy and efficiency. To establish an unbiased baseline that can form a foundation for future developments of toe-based interaction technologies, we opted to study unrestricted interaction without the influence of footwear. Based on the analysis of previous work in the field of biomechanics of toe movements, we formulated the following hypotheses to guide our investigation:

\begin{description}
	\item[\hypothesis{1}] The accuracy of unrestricted voluntary toe movements is sufficient to distinguish targets.
	\item[\hypothesis{2}] Hallux and lateral toes can be used separately for interaction while not restricted.
	\item[\hypothesis{3}] While standing, the interaction of unrestricted toes in flexion direction results in lower performance due to the resistance of the floor.
	\item[\hypothesis{4}] Sitting allows for more accurate, efficient, and comfortable interaction of unrestricted toes compared to standing.
\end{description}

\begin{figure*}[ht!]
	\subfloat[Optical tracking system\label{fig:tictactoes/studysetup:tracking}]
	{\includegraphics[width=.32\linewidth]{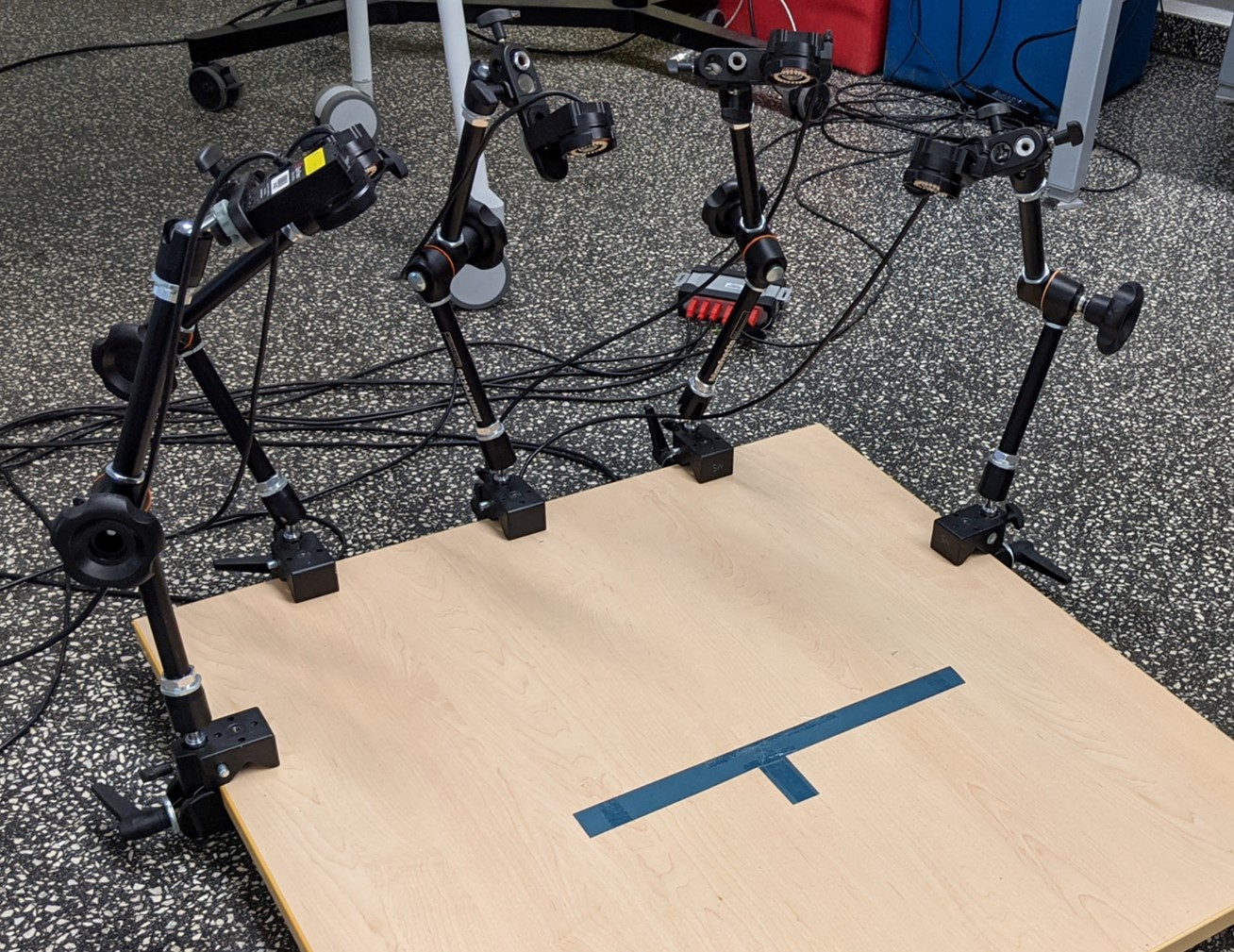}}\hfill
	\subfloat[Markers attached to \toesHallux{} and \nth{3} toe\label{fig:tictactoes/studysetup:markers}]
	{\includegraphics[width=.32\linewidth]{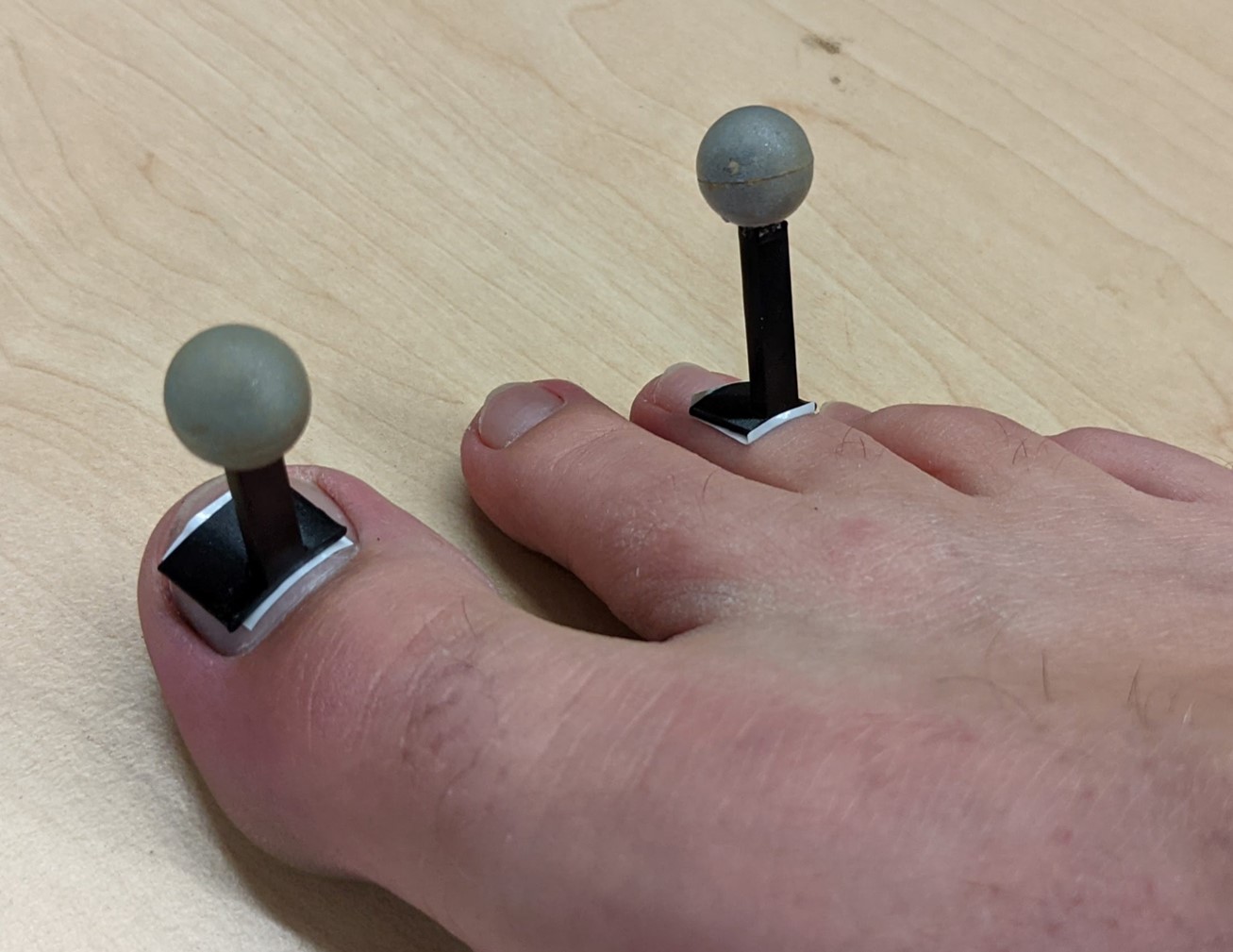}}\hfill
	\subfloat[Study client application\label{fig:tictactoes/studysetup:studyclient}]
	{\includegraphics[width=.32\linewidth]{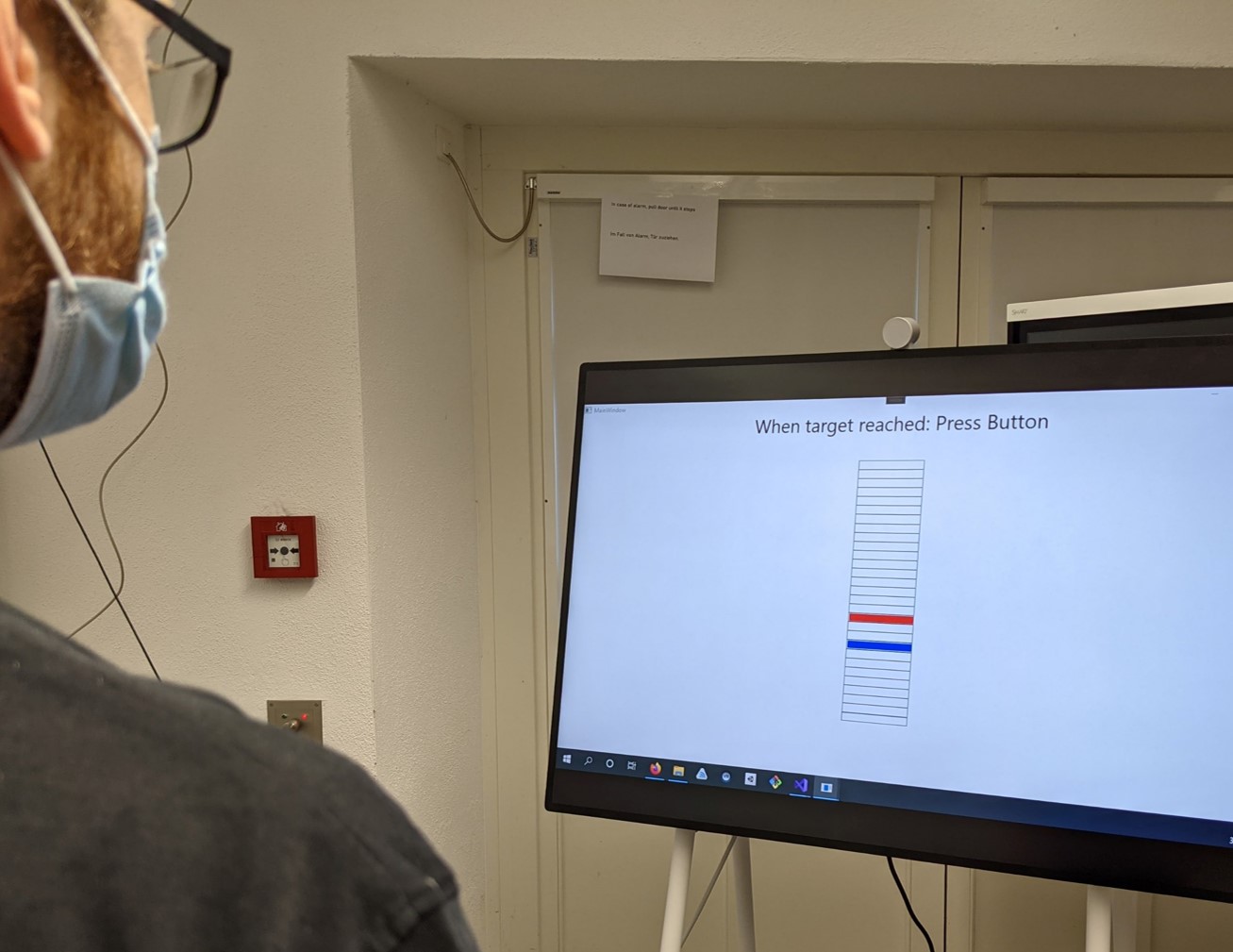}}
	\caption{We used an optical tracking system to (a) track the position and orientation of the participants' \toesHallux{} and 3\textsuperscript{rd} toe using retro-reflective markers attached to 3d-printed extension tubes (b). The study client application visualized the task and visual feedback indicating the current toe deflection angle on a nearby display (c).}
	\Description[An illustration of the study setup.]{An illustration of the study setup. From left to right, the figure depicts a) the optical tracking system, b) feet with attached retro-reflective markers on the hallux and 3rd toe, c) a standing person looking at a display that shows a task as depicted in figure 3.}
\end{figure*}

\subsection{Design and Task}

We designed a controlled experiment in which participants interacted with a system using toe movements of their dominant foot. We visualized participants' toe movements abstractly as a highlighted box on a scale on a nearby display (see fig. \ref{fig:tictactoes/studysetup:tracking}, \ref{fig:tictactoes/task:1}). The participants' first task was to move their toes to a target marked on the scale (see fig. \ref{fig:tictactoes/task:2}, \ref{fig:tictactoes/task:3}) and confirm reaching the target with a button in their hand. As a second task, we asked the participants to keep their toes steady for 3 seconds without visual feedback on their performance after the confirmation click (see fig. \ref{fig:tictactoes/task:4}).

To gain a comprehensive understanding of the potential factors that influence the accuracy, efficiency, and usability of toe-based interfaces, we varied five independent variables (see fig. \ref{fig:tictactoes/factors}):

\begin{description}
	\item[Toes Used for Interaction] Following the inability to move the lateral toes individually, we varied the \factorToes{} variable between \toesAll{}, \toesHallux{}, and \toesLateral{}. In the \toesHallux{} and \toesLateral{} conditions, we mapped the displacement of the respective toe (calibrated to personal maxima and minima) to the scale. For the \toesAll{} condition, we used the mean value between the hallux and the lateral toes.
	\item[Posture] 	We varied the \factorPosture{} between \standing{} and \sitting{}. According to the literature, we excluded walking and other modes of locomotion from the study design because additional voluntary movement of toes during locomotion appears infeasible.
	\item[Direction of Interaction] We varied the \factorDirection{} between \dirUp{} (upwards movement) and \dirDown{} (downward movement). Based on the analysis of toe biomechanics in the literature, we decided against including lateral spreading movements through abduction and adduction because the toe joints function mainly as hinge joints for upward and downward movement.
	\item[Scaling] We varied the subdivision of the interaction range as the \factorScale{} from 4 subdivisions (\scale{4}) over 12 subdivisions (\scale{12}) to 28 subdivisions (\scale{28}). We chose the levels based on informal pre-tests, indicating that these levels provided a good representation across straightforward to tough scales for participants. 
	\item[Target] We varied the \factorTarget{} the participants had to reach. For this, we divided the interaction range into four equal-sized areas and selected the centers of these fields, resulting in targets at \aptLtoX[graphics=no, type=html]{1/8}{\nicefrac{1}{8}}, \aptLtoX[graphics=no, type=html]{3/8}{\nicefrac{3}{8}}, \aptLtoX[graphics=no, type=html]{5/8}{\nicefrac{5}{8}}, and \aptLtoX[graphics=no, type=html]{7/8}{\nicefrac{7}{8}} of the interaction range (\aptLtoX[graphics=no, type=html]{1/8-target}{\target{1}{8}} -- \aptLtoX[graphics=no, type=html]{7/8-target}{\target{7}{8}}). As the levels of \factorScale{} were multiples of 4, each level of \factorTarget{} was in the center of a subdivision for each level of \factorScale{}, making them comparable for later analyses.
\end{description}

We varied these independent variables in a repeated measures design, totaling $3 \cdot 2 \cdot 2 \cdot 3 \cdot 4=144$~conditions. Further, we included 3 repetitions for each condition into the design, resulting in a total of $144 \cdot 3 = 432$~individual trials per participant. We counterbalanced the order of \factorToes{}, \factorDirection{}, and \factorScale{} in a balanced Latin square design with $3 \cdot 2 \cdot 3 = 18$~levels to avoid learning effects. We excluded the \factorPosture{} from the balanced Latin square to avoid fatigue from immediately successive conditions. Instead, we alternated the \factorPosture{} for each participant (i.e., half of the participants performed all \standing{} conditions at the even-numbered positions of the sequence and the \sitting{} conditions at the odd-numbered positions, and vice versa for the other half). For practical reasons, we conducted all levels of \factorTarget{} together with consecutive repetitions for each combination of \factorToes{}~x~\factorDirection{}~x~\factorPosture{}~x~\factorScale{}. Within these blocks, we randomized the levels of \factorTarget{} and the repetitions collectively, resulting in $4 \cdot 3 = 12$ trials per block.

\subsection{Study Setup and Apparatus}

To provide a reliable baseline for future toe-based interfaces, we decided to use an optical tracking system (Optitrack) for accurate and reliable tracking (see fig. \ref{fig:tictactoes/studysetup:tracking}). For this, we mounted the cameras of the tracking system close to the floor and attached retro-reflective markers to the hallux and the \aptLtoX{3\textsuperscript{rd}}{\nth{3}} toe of the participants (see fig. \ref{fig:tictactoes/studysetup:markers}). In addition, we used about \SI{3}{\cm} long 3d-printed extension tubes between the toes and the retro-reflective markers to extend the movement path of the markers to improve tracking performance. 

We calibrated the system to the participants' movement range. Due to the complex bone and muscle apparatus, the displacement of the toes cannot be expressed only as a change in height, but describes a circular path. Therefore, we traced the movement of the markers in the calibration phase from the starting position (flat foot on the floor with relaxed toes) to full flexion or extension, respectively. During the experiment, the system used this calibration to map the deflection of the foot to a percentage value. To distinguish the stationary starting position from directed movements, we considered \SI{\pm 2}{\percent} around the starting position as a neutral zone.  Therefore, the interaction range considered in the experiment refers to the remaining \SI{98}{\percent} each in \dirUp{} and \dirDown{} direction.

Further, we implemented a study client application that allowed us to set the task from a desktop PC. Using an external monitor, the study client visualized the current \factorScale{} and feedback about the foot movement and the current task to the participants (see fig. \ref{fig:tictactoes/studysetup:studyclient}). The participants used a clicker connected via Bluetooth to signal the completion of a task to the system.

For each trial, we logged the following dependent variables: 

\begin{description}
	\item[Success Rate] as the rate of trials where the target was successfully selected.
	\item[Task-Completion Time] as the time between showing the target to the participant and confirming the arrival.
	\item[Number of Crossings] as the number of times, the target field was crossed before confirmation. A crossing is a movement from the start position beyond the target field and vice versa, i.e., to generate a crossing, the entire field must be fully traversed to avoid flickering of the data at the borders of the field.
	\item[Involuntary Toe Movement] as the movement of the inactive toes for the levels \toesHallux{} and \toesLateral{} of the \factorToes{} in percent of the interaction range.
	\item[Holding Error] the maximum movement within \SI{3}{\second} after the confirmation in percent of the interaction range.
\end{description}

\subsection{Procedure}

After welcoming the participants, we introduced them to the concept and asked them to fill out a consent form and a demographic questionnaire. We used the method proposed by Chapman et al. \cite{Chapman1987} to identify the dominant foot. Then, we attached the retro-reflective markers on the toes of the dominant foot using double-sided adhesive tape and calibrated the system to the participants' toes by recording the maximum displacement in the \dirUp{} and \dirDown{} direction. We recalibrated the starting position before each block to avoid errors due to shifts. The calibration process took 5s on average. To avoid learning effects during the experiment, we then asked the participants to freely explore the system by moving their toes while observing the output on the screen. Participants were free to explore the system as long as they wanted until they felt familiar with and comfortable using it. The exploration phase took around \SI{5}{\minute} on average.

After the free exploration phase, we told the participant about the combination of \factorToes{}~x~\factorDirection{}~x~\factorPosture{}~x~\factorScale{} of the first block. After the participant took the correct \factorPosture{}, we started the system, and the participant was presented with the \factorScale{} of the current condition on the screen. Once ready and in starting position (dominant foot placed flat on the ground), the participant used the clicker to start the first trial. After the click, the system highlighted the target on the screen (see fig. \ref{fig:tictactoes/studysetup:studyclient}) in red and the current position of the toe in blue. Then, the participants moved their toes to the target and confirmed the reach using the clicker. After that, the system started a \si{3s} countdown for the second task and informed the participant to hold their toes still. After completing the countdown, the system told the participant to return to the start position. Once ready again, the participants started the subsequent trial using the clicker. We instructed the participants to focus on accuracy instead of speed. If a goal felt impossible, we instructed participants to achieve it as best they could and confirm that position. 

After all trials of a block, participants answered questions regarding their experiences of the last \factorToes{}~x~\factorDirection{}~x~\factorPosture{}~x~\factorScale{} combination on a 7-point Likert scale (1: strongly disagree, 7: strongly agree) using a tablet. To avoid any influence from fatigue, we instructed the participants to take breaks as needed before starting the subsequent trial. In addition, we enforced a minimum 2-minute break before the next block. During this break, the participants gave further qualitative feedback in a semi-structured interview. Each experiment took about 100 minutes per participant. Our institutional ethics board reviewed and approved the study design.

All participants and the investigator were vaccinated or tested on the same day. Only the investigator and the participant were in the room at any given time. The investigator and participants wore medical face masks throughout the experiment. We disinfected the study setup and all touched surfaces and ventilated the room for 30 minutes. 

\subsection{Participants}

We recruited 18 participants (10 identified as male, 8 as female) from our university's mailing list. The participants were aged between 22 and 31
($\mu = 26$, $\sigma = 2.28$) with 3 left-footed. Four of our participants reported having a diagnosed foot condition (P2, P14: pes valgus, P11, P12: pes planus). However, the participants did not have any problems conducting the experiment. All participants took part in the study voluntarily and without financial compensation. We provided drinks and snacks during and after the study.

\subsection{Analysis}

We analyzed the recorded data using 5-way repeated measures (RM) ANOVAs with the independent variables as discussed above as factors to unveil significant main and interaction effects. We
tested the data for normality using Shapiro-Wilk's test. If the assumption of normality was violated, we performed a non-parametric analysis as described below. In cases where Mauchly's test indicated a violation of the assumption of sphericity, we corrected the tests using the Greenhouse-Geisser method and report the \gge{}. When the RM ANOVA indicated significant results, we used pairwise t-tests with Bonferroni correction for post-hoc analysis. We further report the generalized eta-squared \ges{} as an estimate of the effect size, classified using Cohen's suggestions as small ($>.0099$), medium ($>.0588$), or large ($>.1379$)~\cite{Cohen1988} as proposed by Bakeman~\cite{Bakeman2005}. Further, as an estimate of the influence of the individual factors, we report the \ac{EMM} $\muemm{}$ with 95~\% confidence intervals as proposed by Searle et al. \cite{Searle1980}. For the multi-factorial analysis of non-parametric data, such as the Likert questionnaires, we performed an Aligned Rank Transformation as proposed by Wobbrock et al. \cite{Wobbrock2011} and Elkin et al. \cite{Elkin2021}.

\section{Results}
\label{sec:tictactoes:results}

In the following sections, we report the results of the controlled experiment as described in Section. We present the results structured around the dependent variables and provide a summary of the main results at the beginning of each subsection, followed by detailed statistics.

\subsection{Success Rate}

\begin{figure*}[ht!]
	\includegraphics[width=\linewidth]{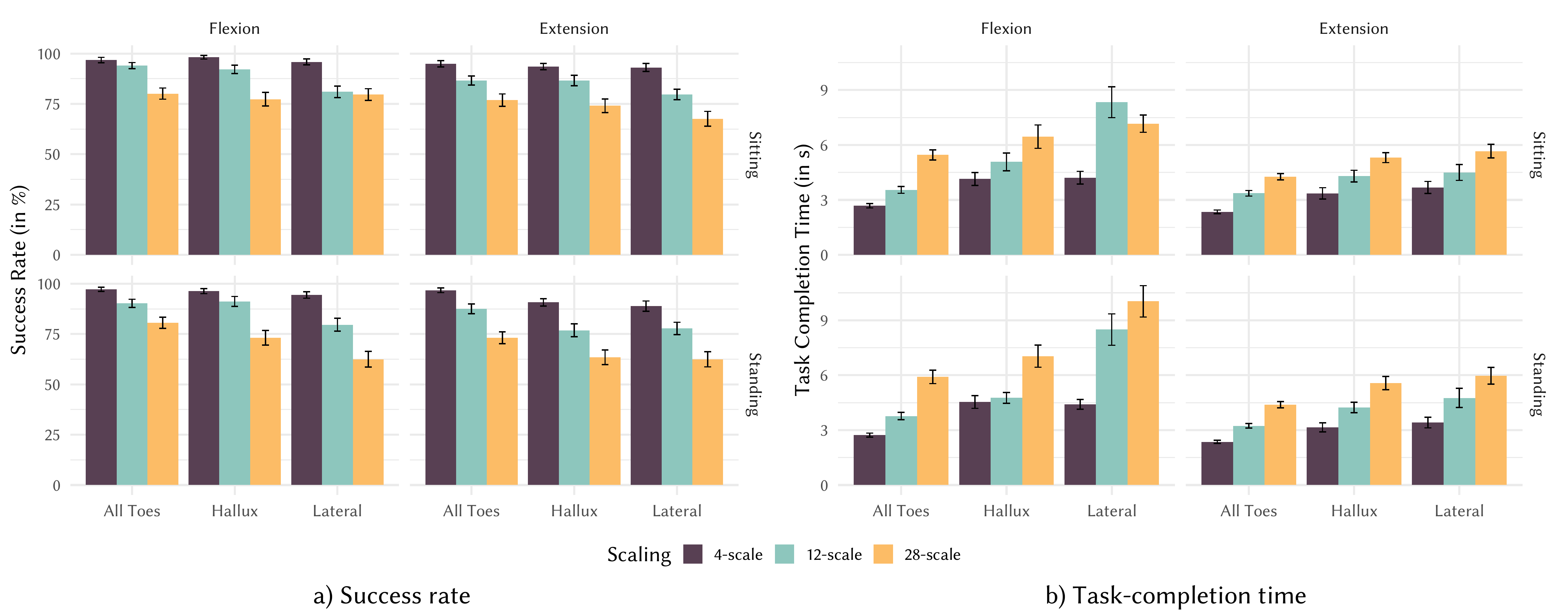}
	\caption{Success rate (a) and task-completion time (b) in the experiment. For ease of reading, data are averaged across the 4 levels of \factorTarget{}. All error bars depict the standard error.}
	\label{fig:tictactoes:success_tct}
	\Description[Bar plots depicting the results of the scperiment.]{Bar plots of the results of the experiment regarding the a) success rate and b) task-completion time.}
\end{figure*}

We measured the success rate when confirming the selection to assess participants' accuracy. We found that \toesLateral{} performed worse compared to \toesAll{} and \toesHallux{}. Further, we found that \sitting{} and \dirDown{} performed better compared to \standing{} and \dirUp{}, respectively. As expected, higher levels of \factorScale{} decreased the accuracy rates. Overall, we measured success rates from \valSi{100}{0}{\percent} (4 times, all \scale{4}) to \valSi{46.30}{30.55}{\percent} (\toesHallux{}, \standing{}, \dirUp{}, \scale{28}, \aptLtoX[graphics=no, type=html]{7/8-target}{\target{7}{8}}), see fig. \ref{fig:tictactoes:success_tct} (a).

The analysis revealed a significant (\anova{2}{34}{12.72}{<.001}{0.02}) main effect of the \factorToes{} with a small effect size. Post-hoc tests showed significantly lower success rates for \toesLateral{} (\emmCI{80.21}{1.27}{77.65}{82.77}{\%}) compared to \toesAll{} (\emmCI{87.89}{1.27}{85.33}{90.45}{\%}, $p<.001$) and \toesHallux{} (\emmCI{84.45}{1.27}{81.89}{87.01}{\%}, $p<.05$).

Further, we found a significant (\anova{1}{17}{23.16}{<.001}{0.01}) main effect of the \factorPosture{} with a small effect size. Post-hoc tests indicated significantly ($p<.001$) higher success rates for \sitting{} (\emmCI{85.98}{0.99}{83.93}{88.03}{\%}) compared to \standing{} (\emmCI{82.38}{0.99}{80.33}{84.43}{\%}).

Also, the analysis indicated a significant (\anova{1}{17}{15.29}{<.01}{0.01}) main effect of the \factorDirection{} with a small effect size. Post-hoc tests confirmed significantly ($p<.01$) higher success rates for \dirDown{} (\emmCI{86.68}{1.12}{84.40}{88.96}{\%}) compared to \dirUp{} (\emmCI{81.69}{1.12}{79.41}{83.97}{\%}).

As a last main effect, the analysis showed a significant (\anovaCor{1.15}{19.54}{113.39}{<.001}{0.57}{0.15}) influence of the \factorScale{} with a large effect size. We found falling success rates for higher levels of \factorScale{}, ranging from \emmCI{94.71}{1.25}{92.19}{97.24}{\%} for \scale{4} to \emmCI{72.57}{1.25}{70.05}{75.09}{\%} for \scale{28}. Post-hoc tests confirmed significant (all $p<.001$) differences between all groups.

Further, we found a significant (\anova{2}{34}{6.46}{<.01}{0.01}) interaction effect between \factorPosture{} and \factorScale{} with a small effect size. While the success rate for the \scale{4} conditions was comparable ($p>.05$) between \standing{} (\emmCI{94.06}{1.38}{91.30}{96.82}{\%}) and \sitting{} (\emmCI{95.37}{1.38}{92.61}{98.13}{\%}), the differences increased with higher levels of scale to \scale{28} (\standing{}: \emmCI{69.21}{1.38}{66.45}{71.97}{\%}, \sitting{}: \emmCI{75.93}{1.38}{73.17}{78.69}{\%}, $p<.001$).

Finally, we found a significant (\anova{3}{51}{5.17}{<.01}{0.01}) interaction effect between \factorDirection{} and \factorTarget{} with a small effect size. The success rates were comparable between \dirUp{} and \dirDown{} from \aptLtoX[graphics=no, type=html]{1/8-target}{\target{1}{8}} to \aptLtoX[graphics=no, type=html]{5/8-target}{\target{5}{8}} with differences between 1\% and 4.5\% in the \acp{EMM} (all $p>.05$). Interestingly, for the \aptLtoX[graphics=no, type=html]{7/8-target}{\target{7}{8}} conditions, however, we found the highest (\dirDown{}: \emmCI{88.68}{1.64}{85.44}{91.93}{\%}) and lowest (\dirUp{}: \emmCI{76.75}{1.64}{73.50}{80.00}{\%}) success rates of all \factorDirection{} and \factorTarget{} combinations ($p<.001$).

\subsection{Task-Completion Time}

As a measure of efficiency, we analyzed the \ac{TCT} as the time between the display of the target and the click to confirm reaching it. We found that \toesAll{}, \sitting{}, and \dirUp{} led to significantly faster \acp{TCT} compared to the respective other levels. As expected, higher levels of \factorScale{} also increased the \ac{TCT}. We found values ranging from \valSi{2.17}{0.56}{s} (\toesAll{}, \standing{}, \dirUp{}, \scale{4}, \aptLtoX[graphics=no, type=html]{3/8-target}{\target{3}{8}}) to \valSi{11.15}{5.8}{s} (\toesLateral{}, \standing{}, \dirDown{}, \scale{28}, \aptLtoX[graphics=no, type=html]{3/8-target}{\target{3}{8}}), see fig. \ref{fig:tictactoes:success_tct} (b).

The analysis showed a significant (\anovaCor{1.39}{23.66}{23.72}{<.001}{0.70}{0.07}) main effect of the \factorToes{} with a medium effect size. Post-hoc tests confirmed rising \acp{TCT} from \toesAll{} (\emmCI{3.67}{0.37}{2.91}{4.43}{s}) over \toesHallux{} (\emmCI{4.83}{0.37}{4.07}{5.59}{s}) to \toesLateral{} (\emmCI{5.89}{0.37}{5.12}{6.65}{s}) with significant differences between all groups (\toesHallux{} - \toesLateral{} $p<.01$, other $p<.001$).

We found a significant (\anova{1}{17}{4.46}{<.05}{0.01}) main effect for the \factorPosture{} with a small effect size. Post-hoc tests confirmed a significant ($p<.05$) faster \acp{TCT} for \sitting{} (\emmCI{4.66}{0.33}{3.97}{5.35}{s}) compared to \standing{} (\emmCI{4.93}{0.33}{4.24}{5.62}{s}).

Also, the analysis showed a significant (\anova{1}{17}{45.44}{<.001}{0.04}) main effect for the \factorDirection{} with a small effect size. Post-hoc tests indicated higher \acp{TCT} for \dirDown{} (\emmCI{5.49}{0.34}{4.78}{6.19}{s}) compared to \dirUp{} (\emmCI{4.10}{0.34}{3.40}{4.81}{s}), $p<.001$.

Further, the analysis found a significant (\anova{2}{34}{73.04}{<.001}{0.10}) main effect of the \factorScale{} with a medium effect size. Post-hoc tests confirmed significantly rising \acp{TCT} with higher levels of \factorScale{} (\scale{4}: \emmCI{3.42}{0.35}{2.70}{4.14}{s}, \scale{12}: \emmCI{4.86}{0.35}{4.15}{5.58}{s}, \scale{28}: \emmCI{6.10}{0.35}{5.38}{6.82}{s}, all $p<.001$).

As a last main effect, the analysis indicated a significant (\anovaCor{1.76}{29.90}{13.16}{<.001}{0.59}{0.03}) influence of the \factorTarget{} with a small effect size. Post-hoc tests confirmed significant differences between \aptLtoX[graphics=no, type=html]{7/8-target}{\target{7}{8}} (\emmCI{5.68}{0.35}{4.95}{6.41}{s}) compared to all other levels (all between $\muemm=$~\SI{4.29}{s} and $\muemm=$~\SI{4.75}{s}, $p<.01$ for \aptLtoX[graphics=no, type=html]{5/8-target}{\target{5}{8}}, rest: $p<.001$).

Besides the main effects, the analysis revealed a significant (\anova{2}{34}{3.74}{<.05}{0.01}) interaction effect between \factorPosture{} and \factorScale{} with a small effect size. While the \acp{TCT} were comparable between \standing{} and \sitting{} for \scale{4} and \scale{12} (both within \SI{0.1}{s}), we found a more pronounced difference for \scale{28} (\standing{}: \emmCI{6.48}{0.36}{5.74}{7.23}{s}, \sitting{}: \emmCI{5.72}{0.36}{4.97}{6.47}{s}). This finding is supported by post-hoc tests confirming significant interaction effects between all conditions except for \scale{4} while \sitting{} compared to \scale{4} while \standing{} and \scale{12} while \sitting{} compared to \scale{12} while \standing{}. 

\begin{figure*}[ht!]
	\includegraphics[width=\linewidth]{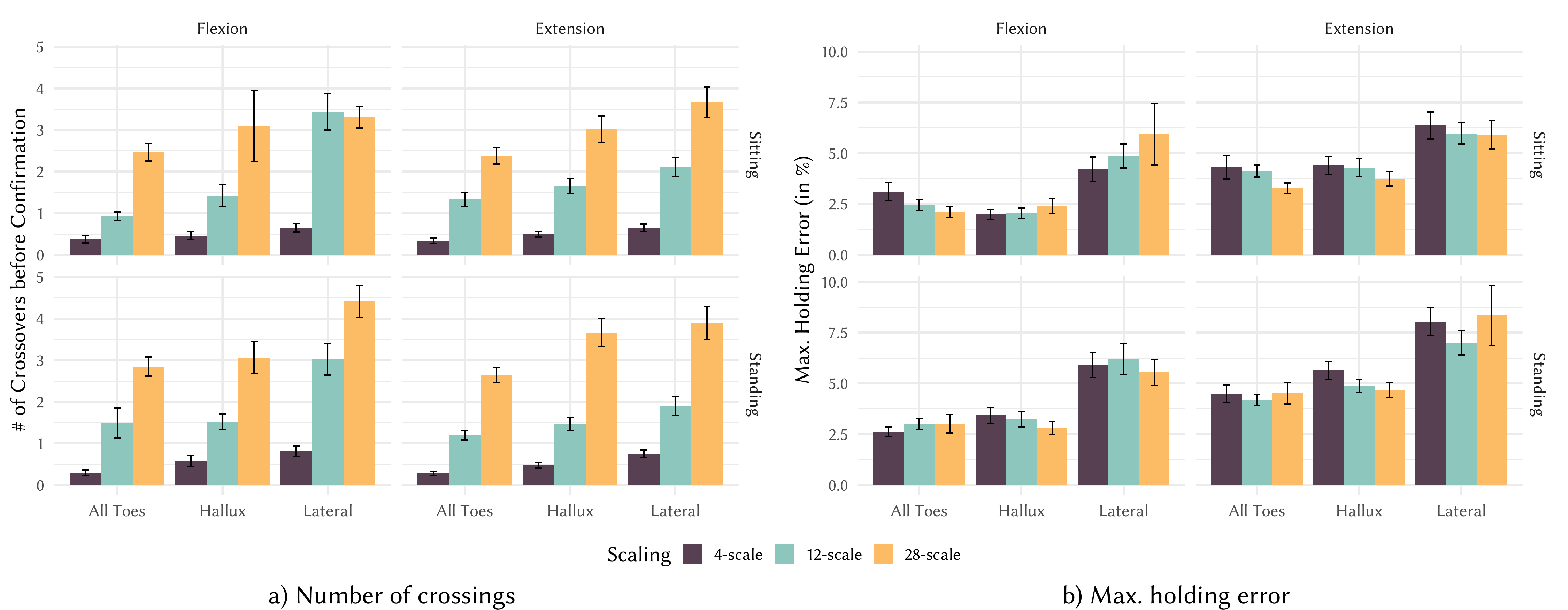}
	\caption{Number of crossings (a) and holding error (b) in the experiment. For ease of reading, data are averaged across the 4 levels of \factorTarget{}. All error bars depict the standard error.}
	\label{fig:tictactoes:crossings_holding}
	\Description[Bar plots depicting the results of the scperiment.]{Bar plots of the results of the experiment regarding the a) number of crossings and b) maximum holding error.}
\end{figure*}

Further, we found a significant (\anova{2}{34}{8.52}{<.01}{0.01}) interaction effect between \factorScale{} and \factorDirection{}. While we could not find significant differences between \dirDown{} and \dirUp{} for \scale{4}, we found significantly higher \acp{TCT} for \dirDown{} movements for both \scale{12} (+\SI{1.6}{s}) and \scale{28} (+\SI{1.8}{s}) (both $p<.001$).

Additionally, we found a significant (\anovaCor{3.23}{54.93}{4.93}{<.01}{0.54}{0.01}) interaction effect between \factorToes{} and \factorTarget{} with a small effect size. While we found a slow and constant, yet not significant, rise of the \acp{TCT} for more distant targets in the \toesAll{} (+\SI{0.5}{s} from \aptLtoX[graphics=no, type=html]{1/8-target}{\target{1}{8}} to \aptLtoX[graphics=no, type=html]{7/8-target}{\target{7}{8}}) and \toesHallux{} (+\SI{0.9}{s}) conditions, the impact on the \acp{TCT} was significant for \toesLateral{} (+\SI{1.7}{s}, $p<.001$).

Finally, we found a significant (\anovaCor{3.00}{51.02}{3.13}{<.05}{0.50}{0.01}) interaction effect between \factorScale{} and \factorTarget{} with a small effect size. More distant targets increased the \acp{TCT} for \scale{28} only slightly (+\SI{0.5}{s} from \aptLtoX[graphics=no, type=html]{1/8-target}{\target{1}{8}} to \aptLtoX[graphics=no, type=html]{7/8-target}{\target{7}{8}}, $p>.05$). For \scale{4} and \scale{12}, however, the rise in \acp{TCT} for more distant targets was more pronounced (\scale{4}: +\SI{1.4}{s}, \scale{12}: +\SI{1.8}{s}, both $p<.01$)

Beyond the two-way interaction effects, we found an interaction effect between \factorToes{}, \factorScale{}, and \factorDirection{} that we omit due to space limitations.

\subsection{Number of Crossings}

We analyzed the number of crossings as another measure of the efficiency of participants. We found higher numbers of crossings for \toesLateral{} compared to \toesAll{} and \toesHallux{}. Further, we found higher numbers of crossings while \standing{} and for the \aptLtoX[graphics=no, type=html]{1/8-target}{\target{1}{8}}. As expected, higher levels of \factorScale{} also led to higher numbers of crossings. Overall, we found numbers of crossings ranging from \val{0.09}{0.25} (\toesAll{}, \standing{}, \dirUp{}, \scale{4}, \aptLtoX[graphics=no, type=html]{7/8-target}{\target{7}{8}}) to \val{5.31}{3.79} (\toesLateral{}, \sitting{}, \dirUp{}, \scale{28}, \aptLtoX[graphics=no, type=html]{1/8-target}{\target{1}{8}}), see fig. \ref{fig:tictactoes:crossings_holding} (a).

The analysis showed a significant (\anova{2}{34}{17.50}{<.001}{0.03}) main effect of the \factorToes{} with a small effect size. Post-hoc tests confirmed significantly higher numbers of crossings for \toesLateral{} (\emmCI{2.39}{0.21}{1.96}{2.81}{}) compared to \toesAll{} (\emmCI{1.38}{0.21}{0.96}{1.80}{}, $p<.001$) and \toesHallux{} (\emmCI{1.74}{0.21}{1.32}{2.17}{}, $p<.01$).

Further, we found a significant (\anova{1}{17}{5.89}{<.05}{0.01}) main effect of the \factorPosture{} with a small effect size. Post-hoc tests confirmed significantly ($p<.05$) lower numbers of crossings while \sitting{} (\emmCI{1.77}{0.18}{1.38}{2.15}{}) compared to \standing{} (\emmCI{1.91}{0.18}{1.52}{2.29}{}).

Also, the analysis revealed a significant (\anovaCor{1.24}{21.13}{77.57}{<.001}{0.62}{0.20}) main effect of the \factorScale{} with a large effect size. Post-hoc tests confirmed significantly rising numbers of crossings for higher levels of \factorScale{} ranging from \emmCI{0.51}{0.22}{0.07}{0.96}{} (\scale{4}) to \emmCI{3.20}{0.22}{2.76}{3.65}{} (\scale{28}), all $p<.005$.

As a last main effect, we found a significant (\anovaCor{2.17}{36.90}{12.61}{<.001}{0.72}{0.03}) influence of the \factorTarget{} with a small effect size. Post-hoc tests revealed significantly higher numbers of crossings for \aptLtoX[graphics=no, type=html]{1/8-target}{\target{1}{8}} (\emmCI{2.42}{0.21}{2.00}{2.84}{}) compared to all other groups ($\muemm{}$ between 1.54 and 1.82, $p<.01$ for \aptLtoX[graphics=no, type=html]{1/8-target}{\target{1}{8}} - \aptLtoX[graphics=no, type=html]{3/8-target}{\target{3}{8}}, rest: $p<.001$)

Besides the main effects, we found a significant (\anova{2}{34}{4.19}{<.05}{0.01}) interaction effect between \factorToes{} and \factorDirection{} with a small effect size. While the numbers of crossings were comparable between \dirDown{} and \dirUp{} for \toesAll{} (difference: 0.04) and \toesHallux{} (difference: 0.11), the difference effect seemed stronger on \toesLateral{} (difference: 0.45). However, post-hoc tests did not support this finding ($p>.05$).

\subsection{Holding Error}

\begin{figure*}[ht!]
	\includegraphics[width=\linewidth]{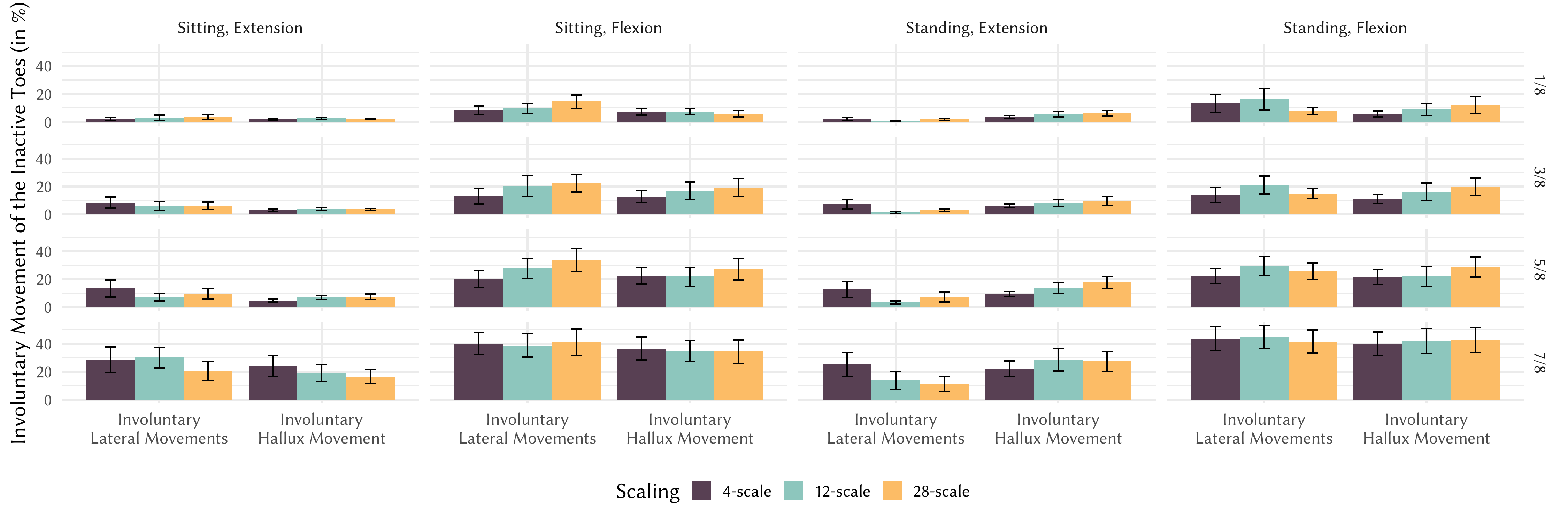}
	\caption{Involuntary movement of the toes that are not active in the respective condition. All error bars depict the standard error.}
	\label{fig:tictactoes:othertoe}
	\Description[Bar plots depicting the results of the scperiment.]{Bar plots of the results of the experiment regarding the involuntary movement of toes.}
\end{figure*}

We analyzed the holding error as the maximum deviation during the hold phase within 3s in percent of the interaction range. We found higher holding errors when using the \toesLateral{} and while \standing{}. Further, we found that \dirUp{} increased the holding error and that \aptLtoX[graphics=no, type=html]{1/8-target}{\target{1}{8}} (in both directions closest to the starting position) showed lower holding errors. Over all conditions, we found holding errors ranging from \valSi{1.27}{1.15}{\percent} (\toesHallux{}, \sitting{}, \dirDown{}, \scale{4}, \aptLtoX[graphics=no, type=html]{3/8-target}{\target{3}{8}}) to \valSi{10.56}{7.6}{\percent} (\toesLateral{}, \standing{}, \dirUp{}, \scale{4}, \aptLtoX[graphics=no, type=html]{3/8-target}{\target{3}{8}}), see fig. \ref{fig:tictactoes:crossings_holding} (b).

The analysis revealed a significant (\anovaCor{1.20}{20.47}{28.84}{<.001}{0.60}{0.07}) main effect of the \factorToes{} with a medium effect size. Post-hoc tests confirmed significantly higher holding errors for \toesLateral{} (\emmCI{6.19}{0.43}{5.30}{7.07}{\%}) compared to \toesAll{} (\emmCI{3.43}{0.43}{2.55}{4.32}{\%}, $p<.001$) and \toesHallux{} (\emmCI{3.63}{0.43}{2.74}{4.51}{\%}, $p<.001$).

Further, the analyses showed a significant (\anova{1}{17}{30.61}{<.001}{0.01}) main effect of the \factorPosture{} with a small effect size. Post-hoc tests revealed significantly ($p<.001$) lower holding errors for \sitting{} (\emmCI{3.97}{0.37}{3.19}{4.76}{\%}) compared to \standing{} (\emmCI{4.86}{0.37}{4.07}{5.64}{\%}).

Also, the analysis showed a significant (\anova{1}{17}{18.79}{<.001}{0.03}) main effect of the \factorDirection{} with a small effect size. Post-hoc tests confirmed significantly ($p<.001$) smaller holding errors for \dirDown{} (\emmCI{3.60}{0.41}{2.76}{4.45}{\%}) compared to \dirUp{} (\emmCI{5.23}{0.41}{4.38}{6.08}{\%}).

As a last main effect, we found a significant (\anovaCor{1.50}{25.56}{3.42}{>.05}{0.50}{0.01}) influence of the \factorTarget{} with a small effect size. Post-hoc tests confirmed significantly lower holding errors for \aptLtoX[graphics=no, type=html]{1/8-target}{\target{1}{8}} (\emmCI{3.95}{0.40}{3.12}{4.79}{\%}) compared to \aptLtoX[graphics=no, type=html]{7/8-target}{\target{7}{8}} (\emmCI{4.79}{0.40}{3.96}{5.62}{\%}), $p<.05$.

In addition, we found a significant (\anovaCor{1.73}{29.38}{10.54}{<.001}{0.58}{0.02}) interaction effect between \factorDirection{} and \factorTarget{}. The holding error was comparable for \aptLtoX[graphics=no, type=html]{1/8-target}{\target{1}{8}} (\dirDown{}: \emmCI{3.99}{0.48}{3.02}{4.96}{\%}, \dirUp{}: \emmCI{3.92}{0.48}{2.95}{4.89}{\%}, $p>.05$). Interestingly, for higher levels of \factorTarget{}, the holding error rose for \dirUp{} (\aptLtoX[graphics=no, type=html]{7/8-target}{\target{7}{8}}: \emmCI{6.39}{0.48}{5.42}{7.36}{\%}) and, at the same time, declined for \dirDown{} (\aptLtoX[graphics=no, type=html]{7/8-target}{\target{7}{8}}: \emmCI{3.19}{0.48}{2.22}{4.16}{\%}), $p<.001$.

Additionally, we found three higher-level interactions that we omit due to space limitations.

\subsection{Involuntary Movement of Inactive Toes}

To gauge the capability for separate voluntary movement of the toes, we measured the movement of the toes not active in the \toesLateral{} and \toesHallux{} conditions. We excluded one participant from the analysis due to technical difficulties with data collection. The measurements  are in \% of the interaction range. We found fewer involuntary toe movements for \dirUp{} compared to \dirDown{}. As expected, we found that the \aptLtoX[graphics=no, type=html]{7/8-target}{\target{7}{8}} (which required the highest extension and flexion) also increased involuntary movement. Overall, we found involuntary movements between \valSi{0.86}{1.28}{\percent} (\toesHallux{}, \standing{}, \dirUp{}, \scale{12}, \aptLtoX[graphics=no, type=html]{1/8-target}{\target{1}{8}}) and \valSi{45.02}{33.35}{\percent} (\toesHallux{}, \standing{}, \dirDown{}, \scale{12}, \aptLtoX[graphics=no, type=html]{7/8-target}{\target{7}{8}}), see fig. \ref{fig:tictactoes:othertoe}.

The analysis revealed a significant (\anova{1}{16}{13.56}{<.01}{0.08}) main effect of the \factorDirection{} with a medium effect size. Post-hoc tests confirmed significantly ($p<.01$) less involuntary toe movement for \dirUp{} (\emmCI{10.12}{2.98}{4.04}{16.21}{\%}) compared to \dirDown{} (\emmCI{23.00}{2.98}{16.92}{29.08}{\%}). 

Also, we found a significant (\anovaCor{1.26}{20.16}{28.84}{<.001}{0.42}{0.16}) main effect of the \factorTarget{} with a large effect size. Post-hoc tests indicated significantly higher involuntary toe movements for \aptLtoX[graphics=no, type=html]{7/8-target}{\target{7}{8}} (\emmCI{31.24}{2.97}{25.21}{37.28}{\%}) compared to all other groups ($p<.001$). Further, post-hoc tests showed significantly higher involuntary toe movement for \aptLtoX[graphics=no, type=html]{5/8-target}{\target{5}{8}} (\emmCI{17.36}{2.97}{11.33}{23.40}{\%}) compared to \aptLtoX[graphics=no, type=html]{1/8-target}{\target{1}{8}} (\emmCI{6.37}{2.97}{0.34}{12.40}{\%}).

Besides, we found a significant (\anovaCor{3.44}{54.98}{3.10}{<.05}{0.57}{0.01}) interaction effect between \factorScale{} and \factorTarget{} with a small effect size. While \scale{28} resulted in the highest involuntary toe movement for \aptLtoX[graphics=no, type=html]{1/8-target}{\target{1}{8}} (\emmCI{6.76}{3.18}{0.35}{13.17}{\%}) to \aptLtoX[graphics=no, type=html]{5/8-target}{\target{5}{8}} (\emmCI{19.66}{3.18}{13.25}{26.07}{\%}), it resulted in the lowest involuntary toe movement for the \aptLtoX[graphics=no, type=html]{7/8-target}{\target{7}{8}} (\emmCI{29.48}{3.18}{23.08}{35.89}{\%}) compared to \scale{4} (\emmCI{32.65}{3.18}{26.24}{39.05}{\%}) and \scale{12} ({\emmCI{31.60}{3.18}{25.19}{38.00}{\%}}). However, post-hoc tests did not confirm this finding (all $p>.05$).

Beyond two-way interaction effects, we found an interaction effect between \factorToes{}, \factorPosture{}, and \factorScale{} that we omit due to space limitations.

\subsection{Questionnaire}
\begin{figure*}[ht!]
\includegraphics[width=\linewidth]{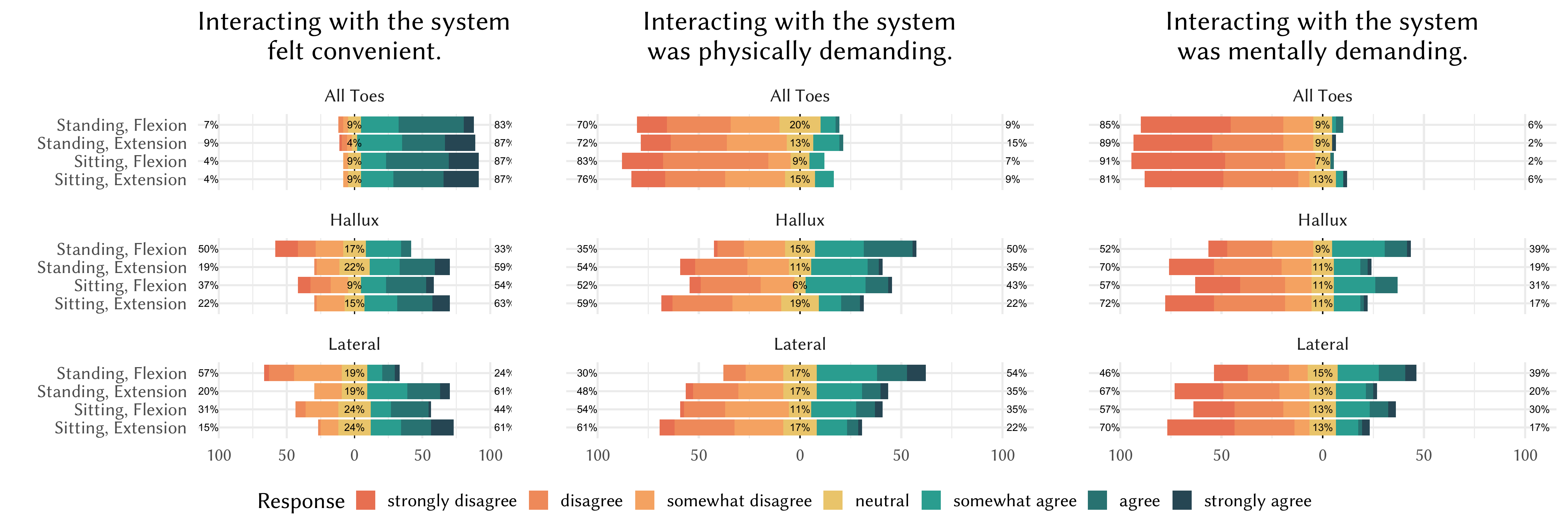}
\caption{The participants’ answers to our four questions in the Likert-questionaires. The plots contain the data of all levels of \factorScale{}.}
\label{fig:tictactoes:likert}
\Description[Plots of the responses of the participants on a 7-point Likert scale.]{Plots of the responses of the participants on a 7-point Likert scale regarding the statements a) Interacting with the system felt convenient, b) Interacting with the system was physically demanding. and c) Interacting with the system was mentally demanding.}
\end{figure*}

We asked participants for subjective feedback on a 7-point Likert scale after each block. The participants, therefore, answered the questions for all targets of the condition together. Figure \ref{fig:tictactoes:likert} depicts all answers from our participants.

\paragraph{Convenience}

We found that participants rated \toesAll{}, \sitting{}, and \dirUp{} better compared to the respective other levels of the factors. Further, participants also rated the highest level of \factorScale{} worse compared to the other levels. 

The analysis revealed a significant (\anova{2}{34}{25.85}{<.001}{0.15}) main effect for the \factorToes{} with a large effect size. Post-hoc tests confirmed significantly higher approval rates for \toesAll{} compared to \toesHallux{} and \toesLateral{} (both $p<.001$). Further, we found a significant (\anova{1}{17}{17.50}{<.001}{0.02}) main effect of the \factorPosture{} with a small effect size. Post-hoc tests showed significantly higher approval ratings for \sitting{} compared to \standing{} ($p<.001$). Also, we found a significant (\anova{1}{17}{11.11}{<.01}{0.06}) main effect of the \factorDirection{} with a medium effect size. Post-ho tests confirmed significantly higher approval rates for \dirUp{} compared to \dirDown{} ($p<.01$). Finally, we found a significant (\anova{2}{34}{6.33}{<.01}{0.01}) main effect for the \factorScale{} with a small effect size. Post-hoc tests showed significantly lower approval rates for \scale{28} compared to both \scale{4} ($p<.01$) and \scale{12} ($p<.05$).

Further, we found a significant (\anova{2}{34}{6.17}{<.01}{0.02}) interaction effect between \factorToes{} and \factorDirection{} with a small effect size. While we could not find significant differences between \dirDown{} and \dirUp{} for \toesAll{} ($p>.05$), we found significantly higher approval for \dirUp{} compared to \dirDown{} for \toesHallux{} and \toesLateral{} (both $p<.001$). Finally, we found a significant (\anova{1}{17}{21.89}{<.001}{0.01}) interaction effect between \factorPosture{} and \factorDirection{} with a small effect size. While we did not see significant differences in the ratings for \dirUp{} while \standing{} and \sitting{}, we found that \dirDown{} resulted in significantly lower ratings while \standing{} compared to \sitting{} ($p<.001$). 

\paragraph{Physical Demand}

We found lower perceived physical demand for \toesAll{} and \sitting{} compared to the other levels of the factors. Further, we found that the highest \factorScale{} increased the perceived physical demand.

We found a significant (\anova{2}{34}{11.45}{<.001}{0.12}) main effect of the \factorToes{} with a medium effect size. Post-hoc tests indicated significantly lower physical demand when using \toesAll{} compared to both \toesHallux{} ($p<.01$) and \toesLateral{} ($p<.001$). Further, the analysis showed a significant (\anova{1}{17}{9.07}{<.01}{0.02}) main effect for the \factorPosture{} with a small effect size. Post hoc test confirmed significantly lower physical demand for \sitting{} compared to \standing{} ($p<.01$). Also, we found a significant (\anova{2}{34}{5.11}{<.05}{0.01}) main effect of the \factorScale{} with a small effect size. Post-hoc tests showed significantly higher physical demand for \scale{28} compared to both \scale{4} and \scale{12} (both $p<.05$).

In addition to the main effects, we found a significant (\anova{2}{34}{6.08}{<.01}{0.02}) interaction effect between \factorToes{} and \factorDirection{} with a small effect size. We found no significant differences between \dirDown{} and \dirUp{} for the \toesAll{} conditions ($p>.05$). For the \toesHallux{} and the \toesLateral{} conditions, however, we found significantly higher physical demand ratings for the respective combinations with \dirDown{} compared to \dirUp{} (\toesHallux{}: $p<.05$, \toesLateral{}: $p<.01$). Finally, we found a significant (\anova{1}{17}{5.75}{<.05}{0.01}) interaction effect between \factorPosture{} and \factorDirection{} with a small effect size. While we found no significant differences between \dirDown{} and \dirUp{} while \sitting{} ($p>.05$), we found a significantly higher physical demand for \dirDown{} compared to \dirUp{} while \standing{} ($p<.01$).

\paragraph{Mental Demand}

We found that \toesAll{}, \sitting{} and \dirUp{} interactions led to a reduced perceived mental demand compared to the respective other levels of the factors. Further, the highest subdivision of \factorScale{} also increased the perceived mental demand.

The analysis showed a significant (\anova{2}{34}{14.24}{<.001}{0.09}) main effect of the \factorToes{} with a medium effect size. Post-hoc tests confirmed significantly lower mental demand for \toesAll{} conditions compared to both \toesHallux{} and \toesLateral{} (both $p<.001$). Further, the analysis indicated a significant (\anova{1}{17}{10.88}{<.01}{0.01}) main effect of the \factorPosture{} with a small effect size. Post-hoc tests confirmed significantly lower mental demand while \sitting{} compared to \standing{} ($p<.01$). Also, we found a significant (\anova{1}{17}{13.57}{<.01}{0.02}) main effect of the \factorDirection{} with a small effect size. Post-hoc tests indicated significantly higher mental demand for \dirDown{} compared to \dirUp{} ($p<.01$). As a last main effect, we found a significant (\anova{2}{34}{9.50}{<.001}{0.02}) influence of the \factorScale{} with a small effect size. Post-hoc tests confirmed a significantly higher mental demand for \scale{28} compared to \scale{4} ($p<.001$).

Finally, we found a significant (\anova{2}{34}{7.22}{<.01}{0.02}) interaction effect between \factorToes{} and \factorDirection{} with a small effect size. We found no significant differences between \dirDown{} and \dirUp{} for \toesAll{}. For \toesHallux{} and \toesLateral{}, however, we found significantly higher mental demand for \dirDown{} compared to \dirUp{} (both $p<.001$).

\subsection{Subjective Feedback}

In general, participants responded positively to the concept of interacting by moving their toes and praised the \pquote{great}{9} work as \pquote{[it is] surprisingly easy to use, similar to indirect control via mouse}{12}. P15 imagined using such a system when interacting with an \nquote{AR Interface [\ldots] you can subtly interact with the system when wearing shoes}.

When asked about their favorite \factorToes{}, there was a clear consensus on the \toesAll{} level (16 participants, \toesHallux{}: 2 participants). Asked for the reasons, participants told us that using all toes felt \mpquote{most pleasant}{P0, P13} and \mpquote{easier}{P2, P5, P10, P15} as \pquote{[it was] most natural}{13}. P5 further described the advantage over \toesHallux{} and \toesLateral{} operation: \nquote{Moving individual toes down was impossible for me, up was okay but still mentally more difficult than using all toes}. P12, among the subgroup who preferred interacting using the \toesHallux{}, stated that this provided him with \nquote{more stability while still [being] flexible [but] only for \dirUp{}}.

Regarding \factorPosture{}, participants showed a slight preference for seated interaction (14 participants, \standing{}: 4 participants). Participants described seated interaction as \pquote{less straining}{3} and that the posture \pquote{[helped me] to keep the balance}{4}. P9 added that he had more \nquote{control [\ldots] while there is no body weight on the foot}.  In contrast, other participants felt that the interaction while standing was more \pquote{controllable}{12}. P5 expressed indecision regarding the preference: \nquote{Even though standing was less comfortable, I felt like I had more control [compared to sitting]}.

For the \factorDirection{}, we found a slight preference for \dirUp{} (13 participants, \dirDown{}: 5 participants) as it felt \mpquote{less straining}{P2, P4, P12} and \mpquote{more comfortable}{P5, P8, P13}. While the responses with a preference for the \dirUp{} direction were more concerned with the pleasantness of the interaction, the group that preferred \dirDown{} mentioned that it felt \mpquote{more precise}{P0, P3}. P5 described the relationship: \nquote{[\dirUp{}] felt more comfortable to use, even though I also felt less accurate that way}. Regarding the two tasks, P4 added: \nquote{When holding, \dirDown{} was more pleasant and when moving, \dirUp{}}. Further, participants made their ratings of \factorDirection{} dependent on \factorPosture{} as \pquote{[flexion] is easier [while] sitting [\ldots] because you do not have to keep your balance and you are [\ldots] more accurate}{13}. 

Concerning the \factorScale{}, we found the highest agreement with the \scale{12} (12 participants, \scale{4}:4 participants, \scale{28}: 1 participant). Participants described this intermediate scaling as a \pquote{good compromise}{3} between \scale{28} where it \pquote{was very difficult to be accurate}{5} and \scale{4} with \pquote{too few options}{3}. The participants who preferred a coarser division as in \scale{4} explained that it was \mpquote{easier}{P2, P4, P10} because \pquote{you do not have to be so precise}{8}.

\section{Discussion and Guidelines}
\label{sec:tictactoes:discussion}

Our results indicate that toe movements can provide a viable input modality for efficient and accurate interactions. We found accuracy rates of up to 100\% (i.e., without a single error across all repetitions and all participants) and interaction times (which included recognizing the target, moving the toes, and confirming the target) well below \SI{2.5}{\second}. In combination, we found the best overall performance using \toesAll{} or \toesHallux{} while \sitting{} for \dirDown{} movements on a \scale{4}. Our results showed that the performance varied considerably across the other levels of the five factors. In this section, we discuss our results in relation to our hypotheses and present guidelines for the future use of toe-based interfaces.

\subsection{Guideline 1: Use a small number of options}

As a fundamental premise, we hypothesized that the accuracy of voluntary toe movements is sufficient to discriminate targets and thus enable purposeful interaction (\hypothesis{1}). To assess this, we investigated different partitions of the interaction space as \factorScale{}. Across all measures, we found a strong influence of the scaling on participants' performance. We observed that increasingly finer subdivisions of the \factorScale{} resulted in a) decreasing accuracy rates as well as b) increasing interaction times. In particular, with the finest subdivision, this effect renders targeted interaction impossible. While the relationship between increased difficulty and decreasing accuracy rates seems intuitive, the second finding that increased difficulty also led to a decrease in efficiency was more surprising. 

We attribute both results to participants having difficulties selecting targets at finer scales (and, thus, smaller target ranges). While this directly affected accuracy rates, it also affected efficiency: We saw participants oscillating around the targets, increasing the interaction times. Higher numbers of crossings for these conditions also support this hypothesis. We found this effect independent of the target's position within the scale.

Therefore, our results support \hypothesis{1} and point to a negative influence on interaction time. Based on these results, we suggest using a \scale{4} for both \dirDown{} and \dirUp{} for most situations. This is in line with the holding errors in the range of $\approx$~\SI{\pm 3}{\percent} around the starting point of the holding phase that we measured for these conditions. This allows users to hold their toes in one of the fields of the scale for longer periods, e.g., to select an item, similar to the concept of dwell-time in eye-gaze interaction~\cite{Qian2017}.

\subsection{Guideline 2: Allow joint movements of the toe groups}

Given the separate muscle groups controlling the \toesHallux{} and \toesLateral{}, we hypothesized that these toe groups could be used separately for interaction. Across all measurements, we found a substantial advantage of joint usage of all toes: Interaction using all toes was significantly faster compared to both \toesHallux{} and \toesLateral{}. Regarding accuracy, we found no significant differences between \toesAll{} and \toesHallux{}, while \toesLateral{} showed a significant drop in success rates.  

However, even these comparable accuracy rates between \toesAll{} and \toesHallux{} do not provide a foundation for independent use: In the analysis of involuntary toe movements, we found large movements of the currently passive toes between \SI{22}{\percent} and \SI{40}{\percent}. However, a future system that aims to interpret users' toe movements would now need to distinguish if a movement by a group of toes was intentional or just a dependency effect on the voluntary movement of the first group. Matching this, our participants communicated clearly that they found independent interaction difficult to impossible. This result supports the findings of Yao et al. \cite{Yao2020}, who found that requiring the user to control the deflections of two toes simultaneously severely degrades operability.

Reflecting on the results, we have found ample evidence that contradicts \hypothesis{2}, indicating that the independent use of the toe groups for different interactions should be considered infeasible. Future systems should therefore use other factors (e.g., direction or finer scales) to increase the number of options while interacting.

\begin{figure*}[ht!]
	\subfloat[App Launcher\label{fig:tictactoes/uc:launcher}]
	{\includegraphics[width=.32\linewidth]{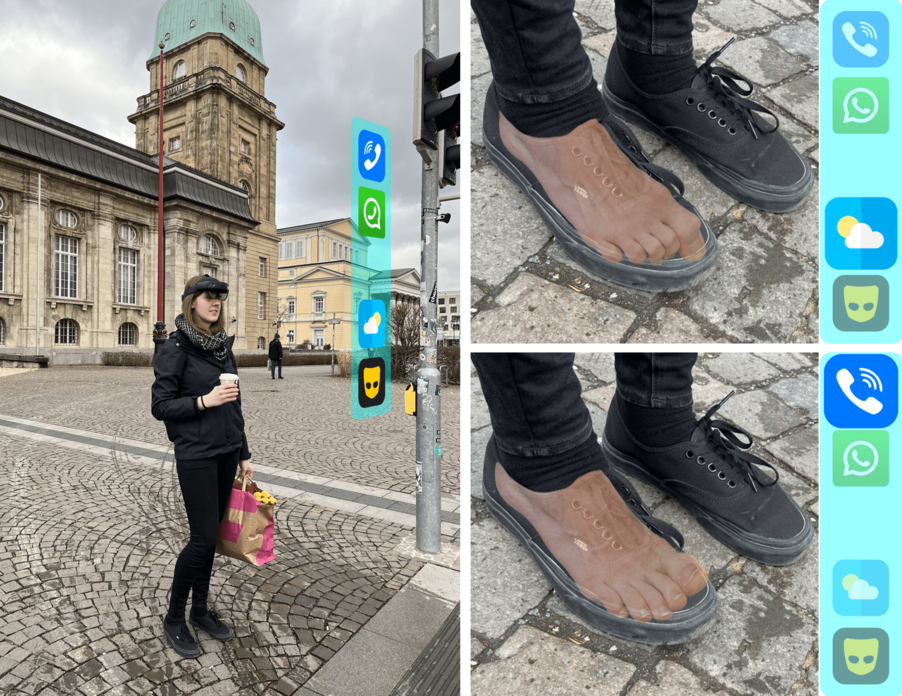}}\hfill
	\subfloat[Music Player\label{fig:tictactoes/uc:music}]
	{\includegraphics[width=.32\linewidth]{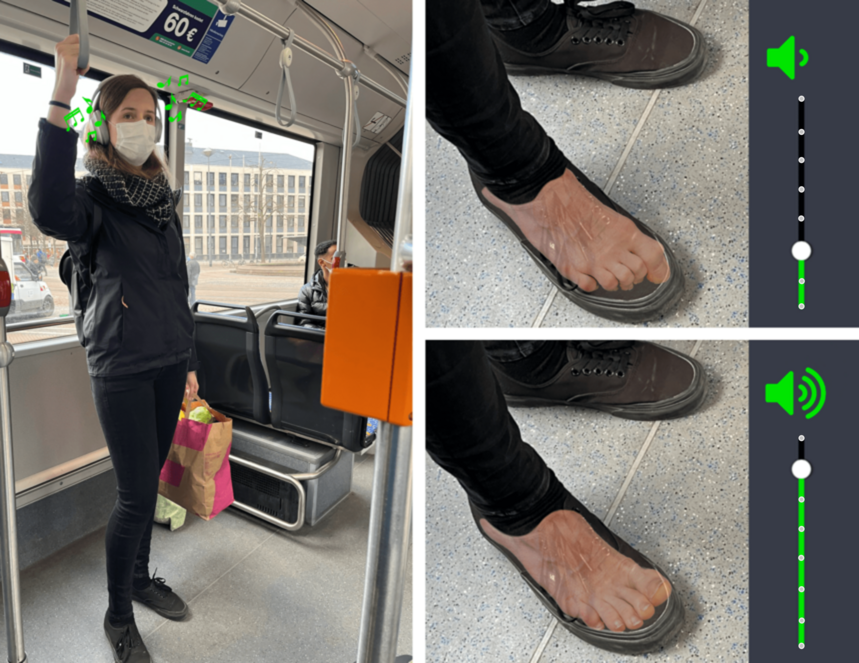}}\hfill
	\subfloat[Kitchen Assistant\label{fig:tictactoes/uc:kitchen}]
	{\includegraphics[width=.32\linewidth]{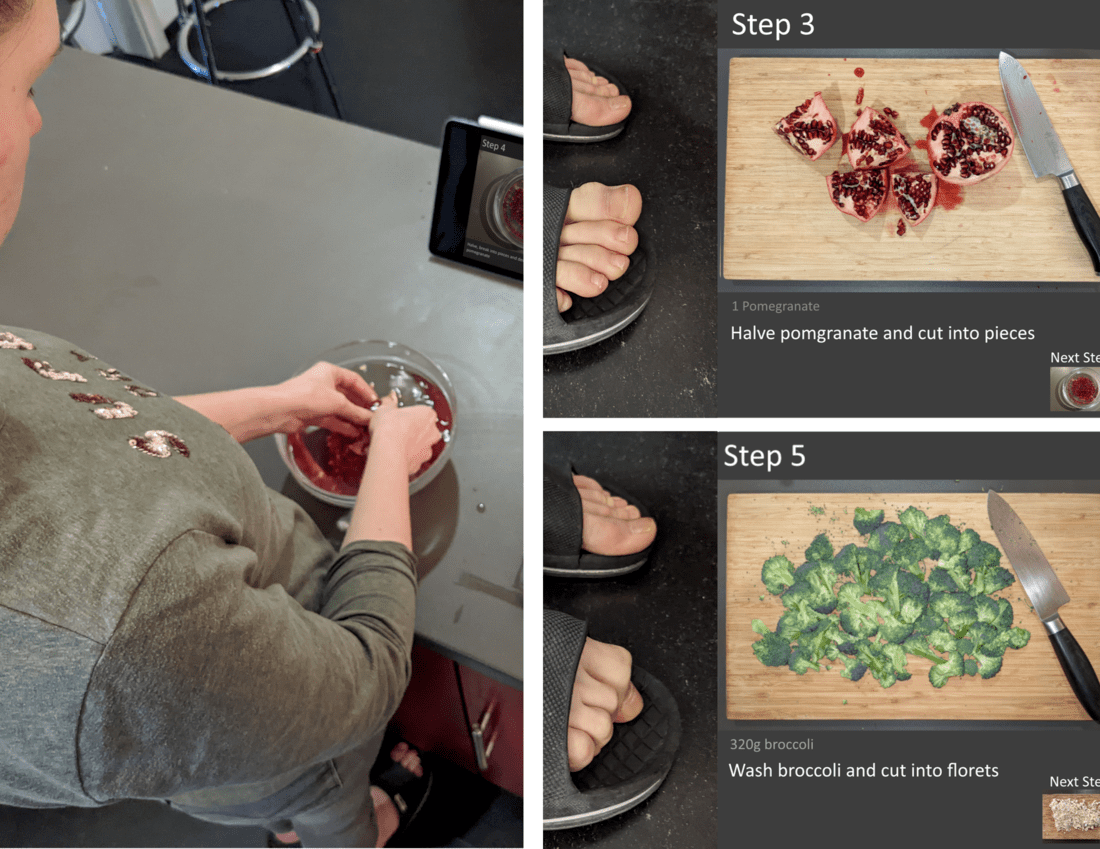}}
	\caption{A set of use cases for a toe-based input modality. The (a) \emph{App Launcher} allows users to select and launch applications on an \ac{HMD}. The (b) \emph{Music Player} allows users to skip songs and change volume without requiring a visual interface. The (c) \emph{Kitchen Assistant} helps users while cooking by allowing them to control the interface on a tablet while their hands are dirty.}
	\Description[Three images illustrating use cases.]{Three images illustrating usecases for toe-based interaction. (a) shows a person wearing an HMD who is waiting at a traffic light with a coffee and a shopping bag in both hands. The person uses foot movements to select an application in the HMD interface. (b) shows a person listening to music in a public transport with headphones and holding on to a handrail to avoid falling over. The person changes the volume by moving his toes. (c) shows a person preparing food in the kitchen while following a receipe on a tablet. The person switches to the next instruction by a toe movement.}
\end{figure*}

\subsection{Guideline 3: Prefer Extension while Standing}

We hypothesized that when standing, the interaction in the flexion direction performs worse than in the extension direction because the toes touching the floor hinders the interaction (\hypothesis{3}). We found significantly higher accuracy rates, faster interaction times, and lower holding errors while \sitting{}. This was also supported by the analysis of the questionnaires, where we found significantly higher ratings for convenience and lower ratings for the physical and mental demand for interacting while \sitting{}.

Looking for reasons for the inferior performance of the participants while \standing{}, we found that especially the combination with \dirDown{} movements yielded a detrimental effect: While the convenience was comparable between \sitting{} and \standing{} for \dirUp{} movements, we found significantly lower approval ratings and higher physical demand for \dirDown{} while \standing{}. We attribute this effect to the permanent muscle tension of the foot and, consequently, the toes while \standing{} to enable us to have a secure and upright stance~\cite{Y2018}. Voluntary movement of the toes interferes with this firm stance. Or the other way around: A firm stance interferes with the voluntary movement of the toes. This is especially relevant for movements in \dirDown{}-direction while \standing{}, where the free movement of the toes is hindered by the floor, so the (fore-) foot has to be slightly lifted. According to our participants' feedback, this destabilized their stance, explaining the deterioration of both quantitative and qualitative performance measures for interaction while \standing{}.

Given these findings, our results provide evidence that supports \hypothesis{3}. Therefore, we suggest focusing on \dirUp{} movements while \standing{}. With \scale{4} using \toesAll{}, \standing{} interactions in \dirUp{}-direction can provide enjoyable interactions with \SI{>95}{\percent} accuracy and \SI{<2.5}{\second} interaction times.

\subsection{Guideline 4: Allow for Additional Options When Sitting}

When seated, the feet are not engaged in maintaining balance, facilitating their use for interaction. Building on this, we hypothesized that sitting allows for more accurate, efficient, and comfortable interaction compared to standing (\hypothesis{4}). Whereas we uncovered a decreasing effect of \dirDown{} movements on the performance of participants while \standing{}, we observed a contrary trend while \sitting{}: While we found accuracy rates well over 90~\% across all conditions for 4 targets, \dirDown{} while \sitting{} proved to be the only combination suitable for keeping comparable accuracy rates for 12 targets. The other conditions, in contrast, showed considerably stronger drops in accuracy rates down to 70-80\%.

As discussed in the last section, maintaining balance while standing puts a substantial strain on the toes, which affects the ability to contribute movement to the interaction. When seated, this effect is eliminated, allowing the toes to demonstrate their full potential. Unleashing this potential, \dirDown{}  movements enabled more fine-grained interactions than \dirUp{} movements. This result came as a surprise, as the musculoskeletal system of the toes supports greater active motion in the \dirUp{} direction than in the \dirDown{} direction~\cite{Creighton1987}. Our results suggest that this larger range of motion could be associated with lower control accuracy.

Based on these findings, we found supporting evidence for \hypothesis{4}. Therefore, we propose to offer additional options while \sitting{}. Thereby, \dirUp{} movements can be used for a coarser selection, while \dirDown{} movements additionally enable a more fine-grained adjustment of options.

\section{Use Cases}
\label{sec:tictactoes:exampleapps}

We acknowledge that toe-based interaction will not replace today's dominant interaction modalities as a stand-alone interface, given the limited variability and expressiveness. However, we are convinced that toe-based input offers great potential as a complementary interaction modality in various situations where today's standard interaction techniques are impractical or unavailable (e.g., when both hands are busy or privacy-sensitive interaction is desired). In the following, we outline example applications illustrating the applicability. 

\paragraph{App Launcher}
The app launcher allows users to scroll through a list to select and open applications on an \ac{HMD}. This use case highlights the possibility of interacting with \ac{AR}~\cite{KnierimHeinSchmidtKosch+2021+49+61} and portable \ac{VR}~\cite{10.1145.3334480.3382920} interfaces in mobile and public scenarios (see Figure~\ref{fig:tictactoes/uc:launcher}) anytime and anywhere. Consequently, toe-based interaction facilitates interaction for users whose hands are occupied or constrained. In addition, toe-based interaction can serve as an intermediate solution for using \ac{AR} in public spaces, providing a smooth transition to social acceptance of \acp{HMD}~\cite{Koelle2017} in public.

\paragraph{Music Player}
\label{sec:tictactoes:exampleapps:musicplayer}
The music player (see Figure~\ref{fig:tictactoes/uc:music}) allows users to skip songs (by extending their toes) and control the volume step by step (by flexing and holding) without using the hands or other input modalities, such as voice-based interaction, which is known to be challenging in noisy environments~\cite{Li2019} and impacts privacy~\cite{Starner2002, Easwara2015}. In contrast, this use case illustrates the appropriateness of toe-based interfaces for private and environment-independent interaction without occupying the environment or compromising user privacy through voice interaction.

\paragraph{Kitchen Assistant}
The kitchen assistant lets users control smart kitchen elements~\cite{10.1145.3316782.3321524, 10.1145.3490632.3490660}, such as a nearby tablet, by moving their toes. This use case spotlights the versatility of our approach to serve as an additional input modality for various devices (see Figure~\ref{fig:tictactoes/uc:kitchen}). Subsequently, toe-based interaction can provide an alternative in situations where the primary input modality of devices is impractical (e.g., when using a tablet with dirty hands).

\section{Limitations and Future Work}
\label{sec:tictactoes:limitations}

We are convinced that the results presented in this paper provide valuable insights into the applicability of toe-based interfaces for interacting with computing systems. However, the design of our experiment and the results impose some limitations and raise questions for future work.

\subsection{External Validity and Real-World Applicability}

In this paper, we aimed to establish a robust baseline of the human capabilities for leveraging voluntary toe movements for interaction. To address the impact of the presented independent variables, we excluded possible confounding factors such as technical challenges of a tracking system or external factors (e.g., restrictions due to footwear). 

We are convinced that the technical challenges of accurate toe-movement tracking will be resolved in the coming years through further technical advances in real-time motion classification using \ac{EMG}~\cite{Qin2020, Karolus2021, 10.1145.3447526.3472027, 10.1145.3546725} or inertial motion tracking~\cite{Lemak2020, Shen2018}. 

Besides the technical limitations, questions remain on how to handle external factors, such as the cognitive demand of the interaction~\cite{kosch2023a}, the shape, or cut of the footwear. Such external factors will influence the freedom of movement of the toes and thus their efficiency and accuracy in the interaction. We see our results as directly applicable to situations without fixed footwear (e.g., barefoot or in socks at home) or to special shoes prepared for toe-based interactions (e.g., sandals or other open-toe shoes). Furthermore, we are convinced that our work provides a solid foundation as an upper bound for further investigations of toe-based interfaces with restrictive footwear. 

\subsection{Confirming a Selection}

In this work, we investigated the influence of several factors on the capabilities of humans to move their toes in a given location. In a potential real-world system, this can be mapped to the selection of an item from a list. However, to use such a selection for interaction, it is necessary to confirm it to avoid unwanted selections during movement.  

As discussed before, a dwell time similar to eye-gaze interaction could be used, in which users keep their toes static in the selected field. We investigated this possibility in our controlled experiment and found holding errors below 3~\% for most conditions, indicating this to be feasible. As a possibly faster option, movements of the toes of the secondary foot or a combination with movements of the whole foot (both the interacting and the secondary foot) could serve as confirmation. Further work is required to conclusively assess the appropriateness of these options to confirm.

\subsection{Continuous Interaction}

The presented experiment focused on the human capabilities to interact with discrete targets on a one-dimensional scale. In potential interaction techniques building on our results, this could be mapped to the sequential selection of discrete options. We chose this design for our experiment to establish a robust baseline and to identify the influence of various factors on the accuracy, efficiency, and user experience while interacting with such a system.

However, we are confident that interaction techniques for toe-based interfaces can also support continuous interactions, such as continuous volume adjustment (cf. use case \emph{Music Player}, section \ref{sec:tictactoes:exampleapps:musicplayer}). For such techniques, the deflection angle of the users' toes could be directly mapped to a cursor or other continuous interface elements. In this field, further work is needed to confirm and quantify the suitability of toe-based interfaces to interact with such continuous interfaces.

\subsection{The Midas Toe Problem}

As with the Midas Touch~\cite{Jacob1995} in eye-gaze interaction and the Midas Tap~\cite{Muller2019} in foot-based interaction, it is a major challenge for a toe-based interaction system to distinguish natural movements in locomotion or balancing while standing from voluntary movements intended to facilitate interaction. As a possible solution, toe-based input systems could be explicitly enabled using a special gesture (e.g., a double grasping gesture of the toes) or a secondary input modality (e.g., a foot movement gesture). As another possibility, toe-based input could be automatically enabled while sitting or standing while being disabled during locomotion using gait detection~\cite{Derawi2010, Ullrich2020}, helping to prevent unintentional interaction.

\subsection{Toe-Based Interactions Without Visual Feedback}

In this work, we investigated toe-based interactions with visual feedback. We chose this approach as our main interest was to gain a robust baseline of our toes' interaction possibilities. Our results thus enable interaction techniques for hands-free interaction with visually-driven interfaces such as \acp{HMD} or nearby screens. 

We acknowledge that visual feedback is not available in every situation. For such situations, other feedback modalities such as audio or vibro (modifying the pitch or vibration intensity, respectively) or even purely proprioceptive~\cite{Lopes2015} or imaginary~\cite{Gustafson2010} interfaces could be used. While we did not test this in the experiment, we are confident that our results in terms of accuracy in interaction and, in particular, holding on to a specific position (done without visual feedback) can provide a solid baseline beyond visual feedback.

\section{Conclusion}
\label{sec:tictactoes:conclusion}

In this work, we explored the idea of leveraging voluntary movements of the toes as an input modality. For this, we investigated the influence of five factors on the accuracy, efficiency, and user experience of voluntary toe movements in a controlled experiment. Our results indicate that the toes can provide a viable input modality for fast yet precise interaction. Further, we found that the granularity of the \factorScale{}, the \factorToes{}, and the \factorPosture{} exert the most substantial influence on the performance and user experience.

\begin{acks}
	We thank our anonymous reviewers for their valuable comments and suggestions and all participants who took part in our experiment. This work has been funded by the \grantsponsor{h2020}{European Union's Horizon 2020 research and innovation program}{http://ec.europa.eu/programmes/horizon2020/en} under grant agreement No.	\grantnum[https://www.humane-ai.eu/]{h2020}{952026}.
	
\end{acks}

\bibliographystyle{bib/ACM-Reference-Format}
\balance{}

\end{document}